\documentclass[journal]{IEEEtran}

\ifCLASSINFOpdf
\else
\fi

\usepackage{amsmath}
\usepackage{algorithm}
\usepackage{algorithmic}
\usepackage{amssymb}
\usepackage{graphicx}
\usepackage{multirow}
\usepackage{array}
\usepackage{booktabs}
\usepackage{diagbox}
\usepackage[caption=false,font=footnotesize]{subfig}
\usepackage{xcolor}
\usepackage{placeins}
\usepackage[colorlinks=true,linkcolor=blue,citecolor=blue,urlcolor=blue]{hyperref}

\begin{document}

\title{Inheritance of Connectivity: Modeling Higher-Order Dynamic Networks with the Birth-Death Process}

\author{
	  Qin~Li,~\IEEEmembership{Member,~IEEE,}
      Yichao~Yao,
      J\"{u}rgen Kurths,
      Matja\v{z}~Perc,~\IEEEmembership{Member,~IEEE,}
      Minyu~Feng,~\IEEEmembership{Senior Member,~IEEE}

  \thanks{This work was supported by the Natural Science Foundation of Chongqing under Grant No. CSTB2025YITP-QCRCX0007, the Fundamental Research Funds for the Central Universities under Grant Nos. XJ2026004101 and SWU-KT26010, and the Slovenian Research and Innovation Agency under Grant No. P1-0403 (corresponding authors: Matja\v{z} Perc and Minyu Feng).}
  \thanks{Qin Li is with the Business College, Southwest University, Chongqing 402460, China.}
  \thanks{J\"{u}rgen Kurths is with the Potsdam Institute for Climate Impact Research, 14437 Potsdam, Germany, and also with the Department of Physics, Humboldt University of Berlin, 10117 Berlin, Germany.}
  \thanks{Matja\v{z} Perc is with the Faculty of Natural Sciences and Mathematics, University of Maribor, Koro\v{s}ka cesta 160, Maribor, 2000, Slovenia, Community Healthcare Center Dr. Adolf Drolc Maribor, Ulica talcev 9, Maribor, 2000, Slovenia, Department of Physics, Kyung Hee University, 26 Kyungheedae-ro, Seoul, 02447, Dongdaemun-gu, Republic of Korea, and University College, Korea University, 145 Anam-ro, Seoul, 02841, Seongbuk-gu, Republic of Korea (e-mail: matjaz.perc@gmail.com).}
  \thanks{Minyu Feng and Yichao Yao are with the College of Artificial Intelligence, Southwest University, Chongqing 400715, China (e-mail: myfeng@swu.edu.cn).}
  }

%
%

\markboth{IEEE TRANSACTIONS ON SYSTEMS, MAN, AND CYBERNETICS: SYSTEMS}{Li \MakeLowercase{\textit{et al.}}: Modeling Higher-Order Dynamic Networks with the Birth-Death Process}
%



\maketitle

\begin{abstract}
Higher-order networks extend pairwise interactions to represent group relationships and provide a useful framework for modeling complex systems. Understanding their dynamic evolution is important for explaining emergent structural and functional properties. However, constructing dynamic higher-order network models with stable topology remains challenging.
In this paper, we propose a higher-order dynamic network model based on an inheritance-based deletion (IBAD) mechanism, which probabilistically redistributes connections when nodes or hyperedges are removed. The model combines birth-death dynamics with preferential attachment, enabling the analysis of evolving higher-order systems. Inheritance is guided by a Gaussian similarity measure based on the distance between degree histories. We derive stationary population distributions and conditional relations for connection retention and degree evolution.
Simulation results show that the model generates networks with statistically stable node and hyperedge populations and heterogeneous degree distributions. The degree distributions and clustering can be adjusted through the node and hyperedge attachment parameters. Comparisons with alternative mechanisms show that inheritance enhances connectivity relative to direct deletion, while the similarity measure influences degree heterogeneity. Comparisons with empirical networks from ecological and communication systems quantify distributional differences using Jensen-Shannon and Kullback-Leibler divergences and reveal discrepancies in clustering and within-hyperedge node-pair counts. These results provide a framework for studying population stability and structural evolution in dynamic higher-order networks.
\end{abstract}
\IEEEpeerreviewmaketitle

\section{Introduction}
%
%
%
%
\IEEEPARstart{O}{ver} the past few decades, complex networks have become a versatile and powerful mathematical tool for analyzing problems across a wide range of domains \cite{barabasi2016,newman2018}. They have been widely applied to study complex systems in ecology \cite{Bairey2016,Grilli2017}, social science \cite{Mastrandrea2015,Benson2016}, epidemic dynamics \cite{Zhai2025}, cyber-physical systems \cite{Li2024} and finance \cite{Bardoscia2021,Ando2022}. Network models capture structural properties and dynamical processes \cite{majhi2022,millan2025topology}, including numerous models of pairwise interactions \cite{Feng2022,pi2025,feng2018,zeng2023,Feng2024}. These models include simple undirected graphs, unsigned and signed networks \cite{Zhang2025}, scale-free networks \cite{barabasi1999}, models with localized evolution \cite{Li2003,Li2023}, weighted evolutionary networks with community structures \cite{Li2006}, and local-world models that incorporate node deletion dynamics \cite{Gu2008}. However, interactions in real-world systems often extend beyond dyadic relationships and involve multi-node or group-level interactions \cite{Estrada2006,Cencetti2021,Battiston2020}. This has motivated the development of higher-order network models, or hypergraphs, which represent such interactions through hyperedges connecting multiple nodes simultaneously \cite{Battiston2021,Battiston2022,bick2023}.

Recent advances in higher-order network modeling have highlighted the ability of hypergraphs to represent complex group interactions \cite{Bianconi2021}. For example, higher-order structures play an important role in opinion dynamics and epidemic spreading, where group influence and contagion are more accurately captured by multi-node interactions than by pairwise links alone \cite{Iacopini2019,St-Onge2021,Bodo2016,Ferraz2023}. In ecological systems, hypergraphs provide a framework for modeling species interactions in which mutualistic or competitive relationships are inherently multi-species \cite{Golubski2016,Levine2017}. Similarly, in social networks, interactions among groups or communities drive the diffusion of behaviors, trends, and information \cite{Zlatic2009,Ferraz2021}. Higher-order interactions also arise in group synchronization \cite{Chen2024} and hyperedge-based pinning control \cite{Yuan2026}. To support the study of systems with higher-order interactions, several models have been developed, including preferential attachment models with heterogeneous degree distributions \cite{Mulder2018}, the simplicial configuration model and the simplicial activity-driven model \cite{young2017,petri2018}, configuration models of random hypernetworks \cite{Chodrow2020}, and hypergraph models with temporal properties \cite{Cencetti2021}, among others.

Despite substantial progress in hypergraph modeling, understanding the dynamic evolution of higher-order networks remains challenging \cite{Gallo2024,Feng2024hyper,Santoro2023}, particularly when nodes and hyperedges are subject to birth and death processes. Configuration models specify structural ensembles, activity-driven and temporal models describe changing interaction patterns, and preferential attachment models explain growth-induced heterogeneity. Recent models describe hyperedge formation through noisy copying \cite{He2025}, dependencies in time-ordered group events \cite{Lerner2025}, and non-Poissonian activation through Markovian node dynamics \cite{Jo2026}. Birth-death models with heritable connectivity provide a starting point for studying connection replacement in pairwise networks \cite{Feng2022}. For higher-order networks, an important modeling challenge is to describe how group interactions reorganize following both node and hyperedge removal, and how this reorganization affects network structure. Studies of reliability under component failures \cite{Wang2026reliability} and spreading with adaptive individual responses \cite{Li2024virus} further motivate examining structural loss and adaptive reorganization. This requires an explicit rule for replacing lost connections and an analysis of its effects on the evolving topology.

To address these issues, we propose a higher-order dynamic network model based on an inheritance-based deletion (IBAD) mechanism. The model combines independent node and hyperedge birth-death processes with preferential attachment. When a node or hyperedge is removed, its connections may be inherited by surviving individuals of the same type. A Gaussian similarity measure based on degree histories guides the selection of replacement individuals. Inheritance uses a single replacement proposal for each removed connection and permits connection loss when replacement fails. This extends similarity-based replacement to both node and hyperedge turnover and allows its structural effects to be examined alongside population dynamics.

In this paper, we investigate the stationarity of node and hyperedge populations and the topological evolution of higher-order dynamic networks under the IBAD mechanism. The contributions are threefold. First, extending pairwise heritable deletion to higher-order networks, we define a degree-history-based inheritance mechanism for both node and hyperedge deletion, with an explicit single-proposal rule that allows replacement failure. Second, we separate population stationarity from topological evolution and derive inheritance-dependent connection-retention and degree-drift relations that account for historical similarity and connection overlap. Third, we examine the effects of model parameters, compare alternative deletion rules and similarity measures, and assess the model using three empirical networks from ecological and communication systems. These analyses characterize the effects of inheritance on degree distributions, clustering, and connectivity, as well as the limitations of empirical reconstruction. From a systems engineering perspective, this framework supports the modeling and assessment of how component turnover and local replacement rules affect population stability and connectivity in systems with group interactions.

\section{Higher-Order Network Model With the IBAD Mechanism}

We model the higher-order network as a queueing system in which nodes and hyperedges undergo continuous-time birth and death events, with inheritance-based reconnection attempted at deletion. In this paper, node and hyperedge populations are treated as two independent birth-death processes that obey the same modeling rule but may have different parameters. Table~\ref{tab:notation} summarizes the principal notation. Subscripts $n$ and $e$ distinguish node and hyperedge quantities. The dependence on time is omitted from the notation when the context is clear.

\begin{table}[!t]
\centering
\caption{Principal Notation}\label{tab:notation}
\footnotesize\setlength{\tabcolsep}{3pt}\renewcommand{\arraystretch}{1.02}
\begin{tabular}{@{}p{.36\columnwidth}>{\raggedright\arraybackslash}p{.60\columnwidth}@{}}
\hline
Symbol & Meaning\\
\hline
$H=(V,E);\ N,M$ & Hypergraph; node and hyperedge counts.\\
$B_{ie};\ L$ & Connection indicator; total connections.\\
$k_i,s_e;\ p_n(k),p_e(s)$ & Node degree, hyperedge size; distributions.\\
$\lambda_n,\lambda_e;\ \mu_n,\mu_e$ & Global birth rates; per-individual death rates.\\
$\tau_n,\tau_e$ & Mean lifetimes, $\tau_n=1/\mu_n$, $\tau_e=1/\mu_e$.\\
$m_n,m_e;\ a_n,a_e$ & Targets per birth; baseline attractiveness.\\
$\Pi_i(S),\Pi_e(S)$ & Next-target probabilities excluding $S$.\\
$A_i^e,A_e^n$ & Inclusion probabilities at opposite-type birth.\\
$\mathcal T_{ij},\ell_{ij};\ h_{\min}$ & Common history, length; minimum length.\\
$D_{ij}^{2},w_{ij};\ \sigma$ & Mean squared history distance, similarity; bandwidth.\\
$\rho_{ij}$ & Pearson history correlation in comparisons.\\
$\mathcal C_i;\ q^n_{ij},q^e_{fh}$ & Eligible survivors; proposal probabilities.\\
$r^n_{ie},r^e_{if}$ & Connection replacement success probabilities.\\
$K_i^n,K_f^e$ & Expected replacements per deletion.\\
$c_{ij};\ G_{ij}$ & Shared hyperedges; gains to $j$ at $i$'s deletion.\\
$b(H);\ g_i^n,g_e^e$ & Birth connection rate; surviving-individual drifts.\\
$\eta_n,\eta_e;\ \widehat\eta_n,\widehat\eta_e$ & Stationary weighted and measured retention fractions.\\
$X;\ \theta$ & Either population; stationary mean $\lambda/\mu$.\\
$P_{rs}(h),p_r(t),\pi_r$ & Transition, marginal, stationary probabilities.\\
$\gamma_n,\gamma_e$ & Decay parameters; fitted slopes $-\gamma_n,-\gamma_e$.\\
$C;\ {\rm CV}_n,{\rm CV}_e$ & Clustering; degree/size coefficients of variation.\\
$N_{\rm LCC},R_{\rm LCC},\ell_{\rm LCC}$ & Largest-component count, fraction, mean path length.\\
$Q_2$ & Within-hyperedge node pairs, with multiplicity.\\
$D_{\rm KL},D_{\rm JS}$ & KL and JS divergences.\\
\hline
\end{tabular}
\end{table}

We consider a hypergraph $H(t)=(V(t),E(t))$ with $N(t)=|V(t)|$ nodes and $M(t)=|E(t)|$ hyperedges. A connection denotes membership of a node in a hyperedge. The binary incidence matrix $B_{ie}(t)$ records these connections: $B_{ie}=1$ if node $i$ belongs to hyperedge $e$, and zero otherwise. The node hyperdegree and hyperedge size are, respectively,
\begin{equation}
 k_i=\sum_{e\in E}B_{ie},\qquad s_e=\sum_{i\in V}B_{ie},\qquad L=\sum_i k_i=\sum_e s_e.
 \label{eq:incidence}
\end{equation}
Here, $L$ is the number of connections. We denote the node-degree and hyperedge-size distributions by $p_n(k)$ and $p_e(s)$; the plotting labels $hk_n$ and $hk_e$ denote node degree and hyperedge size, respectively. Each node-hyperedge pair has at most one connection. Hyperedges with different identities may have the same members. Isolated nodes and empty hyperedges remain in the system until their own deaths.

Each population follows a continuous-time birth-death process on $\{0,1,\dots\}$ under the following queueing assumptions.

\textbf{Queueing System for Evolving Networks:}
\begin{itemize}
\setlength{\itemsep}{0pt}\setlength{\parsep}{0pt}
\item [i)] \textit{Input Process:} We employ two independent homogeneous Poisson processes to describe the arrivals of nodes and hyperedges. Their parameters $\lambda_n$ and $\lambda_e$ determine the corresponding global birth rates.
\item [ii)] \textit{Service Mechanism:} The lifetimes of nodes and hyperedges are modeled by independent exponential distributions with parameters $\mu_n$ and $\mu_e$, respectively. These are per-individual death rates, giving total death rates $\mu_nN(t)$ and $\mu_eM(t)$. The mean lifetimes are $\tau_n=1/\mu_n$ and $\tau_e=1/\mu_e$.
\item [iii)] \textit{Capacity:} The network has no upper size bound; new individuals enter immediately.
\item [iv)] \textit{Service Discipline:} All nodes and hyperedges begin service immediately upon arrival and have independent lifetimes. No waiting queue or service-order rule is required.
\end{itemize}

Accordingly, the node and hyperedge population processes can each be represented by an $M/M/\infty$ queue, where the first $M$ denotes the Poisson arrival process, the second $M$ denotes exponential lifetimes, and $\infty$ represents infinitely many servers accommodating simultaneous independent lifetimes. Inheritance updates connections while preserving the identities and remaining lifetimes of surviving individuals.

\subsection{Growth and Inheritance-Based Deletion of Nodes and Hyperedges}

\begin{figure*}[!t]
\centering
\includegraphics[scale=0.52]{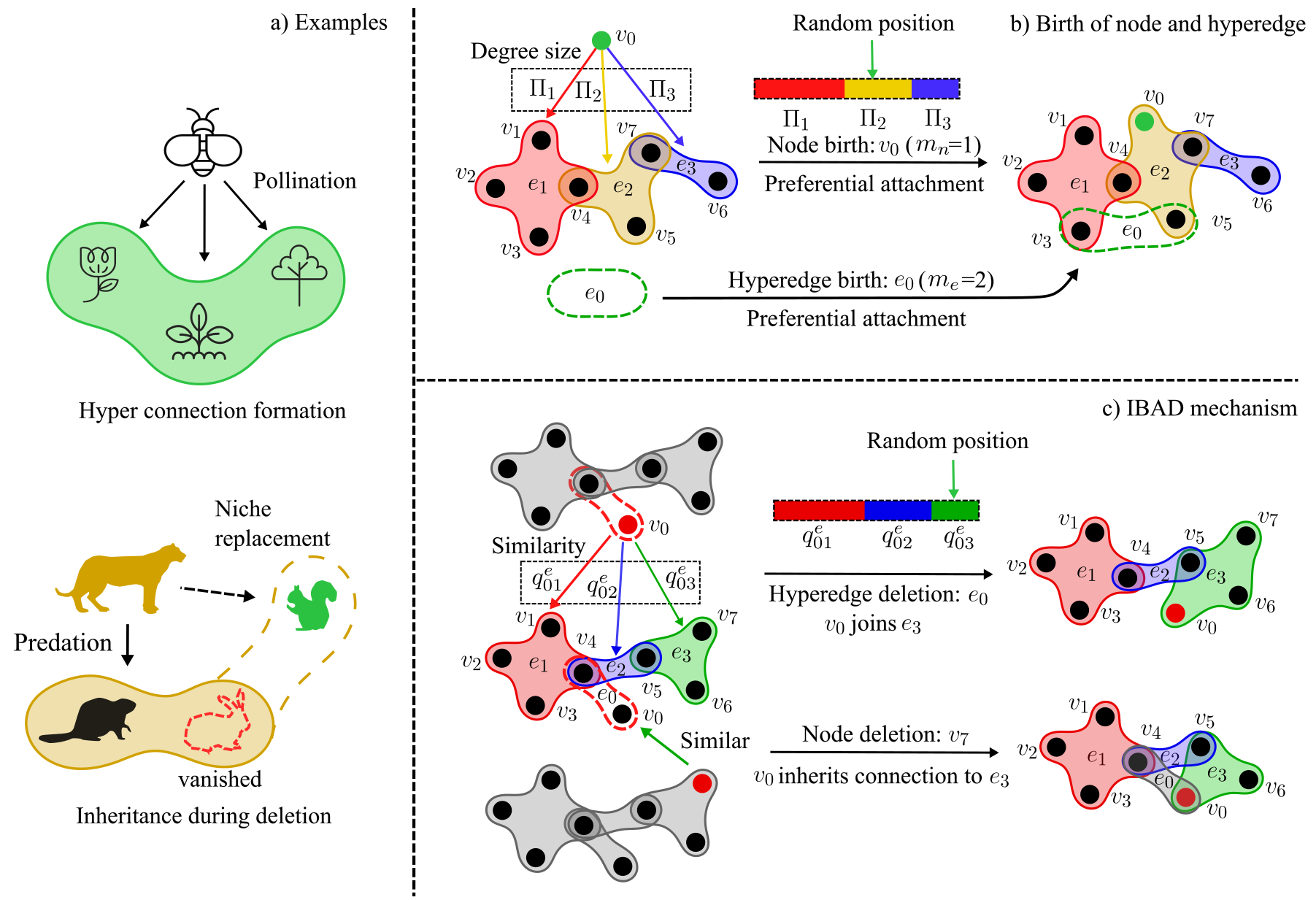}
\caption{\label{fig1}
\textbf{Birth, death, and inheritance-based deletion of nodes and hyperedges.} (a) Examples of higher-order interactions and role replacement. (b) A newborn node $v_0$ selects existing hyperedges by preferential attachment; a newborn hyperedge $e_0$ selects existing nodes analogously. The roulette intervals $\Pi_1,\Pi_2,\Pi_3$ represent the next-target probabilities in \eqref{eq:attachment}. (c) Successful inheritance examples following hyperedge and node deletion. The graphical labels $q^e_{01},q^e_{02},q^e_{03}$ denote the normalized replacement-proposal probabilities in \eqref{eq:proposal}.}
\end{figure*}

\textit{Preferential attachment:} To capture the Matthew effect observed in many real-world networks, we adopt a preferential attachment rule similar to that of Barabási-Albert networks when adding new nodes and hyperedges. A newborn node joins $\min(m_n,M)$ distinct existing hyperedges, whereas a newborn hyperedge selects $\min(m_e,N)$ distinct existing nodes. Targets are sampled sequentially without replacement. If $S$ is the set of already selected targets, the probabilities for the next choice are
\begin{equation}
 \Pi_e(S)=\frac{s_e+a_e}{\sum_{f\notin S}(s_f+a_e)},\qquad
 \Pi_i(S)=\frac{k_i+a_n}{\sum_{j\notin S}(k_j+a_n)}.
 \label{eq:attachment}
\end{equation}
The degrees in these expressions are evaluated before the birth. The positive baseline attractiveness parameters $a_n$ for nodes and $a_e$ for hyperedges allow zero-degree individuals to receive new connections, extending linear preferential attachment \cite{barabasi1999,Dorogovtsev2000}. When the opposite population is empty, the newborn individual has no connection. Figure~\ref{fig1}(b) illustrates preferential attachment at node and hyperedge birth.

\textit{Inheritance-based deletion:} For node and hyperedge removal, we assume exponential lifetimes for both nodes and hyperedges. The IBAD mechanism is then applied at the moment of deletion. In practical systems such as population networks and scientific collaboration networks, individuals may die or retire and are then removed from the network. When a node is removed, each incident hyperedge independently proposes a surviving node for reconnection, weighted by its similarity to the departing node. Similarly, when a hyperedge is removed, each of its nodes independently proposes a surviving hyperedge, weighted by its similarity to the departing hyperedge. The core of IBAD is therefore the definition of similarity between departing and surviving nodes or hyperedges.

For example, in a professional collaboration network, if a key individual leaves a project, the remaining collaborators may seek a new member with similar skills or experience to preserve the project structure. Here, researchers represent nodes and projects represent hyperedges. The model uses degree histories as a structural proxy for role similarity. If the proposed researcher already participates in the project, no additional connection is created, and the proposal is not redrawn. Conversely, when a project closes, its members may seek places in other projects with similar histories of team size, subject to the same single-proposal rule. Figure~\ref{fig1}(a) presents ecological examples of group interactions and role replacement.

\textit{Degree-history similarity:} The degree is a key topological attribute of both nodes and hyperedges, and its history reflects both the level and temporal variation of their connectivity. To quantify similarity, we compare the degree histories of same-type individuals. After each global birth or death event, we append the current degree of every surviving individual to its history. A history begins at the individual's birth. Just before a deletion, two same-type individuals $i$ and $j$ are compared over their common, simultaneously observed history suffix $\mathcal T_{ij}$. Its length $\ell_{ij}$ counts the common post-event degree records. The parameter $h_{\min}$ is the minimum required number of these records, so a candidate is eligible only if $\ell_{ij}\ge h_{\min}$. All observations in the common suffix enter the comparison once the minimum history length is reached. For nodes, define
\begin{equation}
 D_{ij}^{2}=\frac{1}{\ell_{ij}}\sum_{u\in\mathcal T_{ij}}[k_i(u)-k_j(u)]^2,
 \qquad w_{ij}=\exp\!\left(-\frac{D_{ij}^{2}}{2\sigma^2}\right).
 \label{eq:similarity}
\end{equation}
For hyperedges, replace $k$ by $s$. Let $\mathcal C_i$ be the eligible surviving nodes other than $i$. The proposal probabilities are
\begin{equation}
 q^n_{ij}=\frac{w_{ij}}{\sum_{v\in\mathcal C_i}w_{iv}}\quad(j\in\mathcal C_i),
 \label{eq:proposal}
\end{equation}
with zero probability outside this set. If $\mathcal C_i$ is empty, no proposal is made. We analogously write $q^e_{fh}$ for replacement of hyperedge $f$ by $h$.

Equation~\eqref{eq:similarity} uses the squared-exponential, or Gaussian radial-basis, similarity, a standard distance-to-similarity transformation \cite{Rasmussen2006}. Dividing the squared distance by $\ell_{ij}$ puts histories of different lengths on a common mean-square scale. The bandwidth $\sigma$ has degree units and controls tolerance to an absolute difference in historical degree. Equal constant histories have similarity one, whereas constant histories at different levels remain distinguishable. This is useful when role similarity should reflect both degree level and temporal variation. Pearson correlation instead measures standardized covariation, is insensitive to offsets and positive rescaling, and is undefined for constant histories. These properties motivate the distance-based baseline. Pair-specific common histories define positive sampling weights for the inheritance rule.

A larger $\sigma$ makes proposal weights more uniform over eligible candidates; as $\sigma$ tends to zero, the distribution concentrates on candidates with the smallest historical distance. Replacing $k$ by $\log(1+k)$ emphasizes relative differences and compresses differences among high-degree individuals. We compare this alternative and two Pearson-based rules in Section~\ref{sec:comparison}.

\textit{Deletion and single proposals:} For a departing node $i$, compute $q^n_{ij}$ once from the pre-deletion histories. For each $e$ with $B_{ie}=1$, independently draw one $j$ using this same distribution. If $B_{je}=0$, add $(j,e)$; otherwise skip the proposal. Remove all connections of $i$ and then remove $i$. For a departing hyperedge $f$, independently propose one replacement hyperedge $h$ for each connected node, establish only new connections, and finally remove $f$. The proposal distribution covers all history-eligible same-type survivors and remains fixed throughout the deletion event. The success probability of inheritance is therefore determined by the proposal distribution and the candidates' existing connections. Figure~\ref{fig1}(c) illustrates successful replacement connections following hyperedge deletion (upper example) and node deletion (lower example).

\begin{algorithm}[!t]
\caption{Higher-Order Network Modeling With Single-Proposal IBAD}
\label{alg:ibad}
\renewcommand{\algorithmicrequire}{\textbf{Input:}}
\renewcommand{\algorithmicensure}{\textbf{Output:}}
\footnotesize
\begin{algorithmic}[1]
\REQUIRE $H(0)$, event budget $T$, $\lambda_n,\lambda_e,\mu_n,\mu_e$, $m_n,m_e$
\ENSURE $H$ and recorded network statistics
\STATE $H\leftarrow H(0)$; $\mathcal Q\leftarrow[\,]$; $r\leftarrow0$
\STATE InitializeHistories$(H)$
\FORALL{$v\in V$}
\STATE Schedule$(\mathcal Q,\mu_n,d_n,v)$
\ENDFOR
\FORALL{$e\in E$}
\STATE Schedule$(\mathcal Q,\mu_e,d_e,e)$
\ENDFOR
\STATE Schedule$(\mathcal Q,\lambda_n,b_n,\varnothing)$
\STATE Schedule$(\mathcal Q,\lambda_e,b_e,\varnothing)$
\WHILE{$r<T$}
\STATE $(\Delta t,type,x)\leftarrow$ PopMin$(\mathcal Q)$
\STATE DecreaseRemainingTime$(\mathcal Q,\Delta t)$
\IF{$type=b_n$}
\STATE $v\leftarrow$ CreateNode$()$
\STATE $S\leftarrow$ SelectDistinct$(E,\mathrm{BirthWeights}(E),\min(m_n,|E|))$
\STATE ConnectAll$(v,S)$
\STATE Schedule$(\mathcal Q,\mu_n,d_n,v)$
\STATE Schedule$(\mathcal Q,\lambda_n,b_n,\varnothing)$
\ELSIF{$type=b_e$}
\STATE $e\leftarrow$ CreateHyperedge$()$
\STATE $S\leftarrow$ SelectDistinct$(V,\mathrm{BirthWeights}(V),\min(m_e,|V|))$
\STATE ConnectAll$(S,e)$
\STATE Schedule$(\mathcal Q,\mu_e,d_e,e)$
\STATE Schedule$(\mathcal Q,\lambda_e,b_e,\varnothing)$
\ELSIF{$type=d_n$}
\STATE $\mathcal C_x\leftarrow$ Eligible$(V\setminus\{x\},x)$
\STATE $q^n_{x\cdot}\leftarrow$ HistoryProb$(x,\mathcal C_x)$
\FORALL{$e\in E$ such that $B_{xe}=1$}
\IF{$\mathcal C_x\ne\varnothing$}
\STATE $j\leftarrow$ WeightedSelect$(\mathcal C_x,q^n_{x\cdot})$
\IF{$B_{je}=0$}
\STATE Connect$(j,e)$
\ENDIF
\ENDIF
\ENDFOR
\STATE RemoveNode$(x)$
\ELSE
\STATE $\mathcal C_x\leftarrow$ Eligible$(E\setminus\{x\},x)$
\STATE $q^e_{x\cdot}\leftarrow$ HistoryProb$(x,\mathcal C_x)$
\FORALL{$i\in V$ such that $B_{ix}=1$}
\IF{$\mathcal C_x\ne\varnothing$}
\STATE $f\leftarrow$ WeightedSelect$(\mathcal C_x,q^e_{x\cdot})$
\IF{$B_{if}=0$}
\STATE Connect$(i,f)$
\ENDIF
\ENDIF
\ENDFOR
\STATE RemoveHyperedge$(x)$
\ENDIF
\STATE AppendCurrentDegrees$(H)$; Record$(H,r+1)$
\STATE $r\leftarrow r+1$
\ENDWHILE
\RETURN $H$
\end{algorithmic}
\end{algorithm}

In Algorithm~\ref{alg:ibad}, $\mathcal Q$ is the remaining-time event queue, $T$ is the event budget, $r$ counts completed events, and $\Delta t$ is the time to the next event. The symbols $b_n,b_e,d_n,d_e$ denote node birth, hyperedge birth, node death, and hyperedge death, respectively. Schedule$(\mathcal Q,\rho,z,x)$ inserts $(\mathrm{GenerateExp}(\rho),z,x)$ into the remaining-time queue $\mathcal Q$, with exponential rate $\rho$, scheduled event type $z$, and individual identifier $x$. BirthWeights computes the preferential-attachment weights in \eqref{eq:attachment}, SelectDistinct samples distinct targets using these weights, and ConnectAll adds the resulting connections. Eligible identifies history-eligible survivors. HistoryProb calculates the historical similarities in \eqref{eq:similarity} and normalizes them by \eqref{eq:proposal} once before the connection loop; for an empty set it returns an empty vector. Each WeightedSelect call is independent, and a duplicate proposal is skipped without redrawing. Creation starts an empty history, AppendCurrentDegrees records the post-event degrees, and removal deletes the individual, its original connections, and its history.

The total instantaneous event rate is $\lambda_n+\lambda_e+\mu_nN+\mu_eM$. Independent arrival and lifetime clocks give the corresponding continuous-time process.

\subsection{Stationary Population Distributions}
\label{sec:population}

Let $X(t)$ represent either population, with global arrival rate $\lambda>0$ and per-individual death rate $\mu>0$. Distinguish the transition probability $P_{rs}(h)=\Pr\{X(t+h)=s\mid X(t)=r\}$ from the state probability $p_r(t)=\Pr\{X(t)=r\}$. For fixed $r$ and $h\to0^+$,
\begin{equation}
 \begin{aligned}
 &P_{r,r+1}(h)=\lambda h+o(h),\\
 &P_{r,r-1}(h)=r\mu h+o(h),\\
 &P_{r,r}(h)=1-(\lambda+r\mu)h+o(h).
 \end{aligned}
 \label{eq:transitions}
\end{equation}
Other transitions have probability $o(h)$, and the downward transition is absent at $r=0$. Conditioning on the previous state gives the forward equations
\begin{equation}
 \frac{d p_r(t)}{dt}=\lambda p_{r-1}(t)+(r+1)\mu p_{r+1}(t)-(\lambda+r\mu)p_r(t),
 \label{eq:forward}
\end{equation}
where $p_{-1}(t)=0$.

\textit{Proposition 1:} The stationary distribution is Poisson with mean $\theta=\lambda/\mu$:
\begin{equation}
 \pi_r=e^{-\theta}\frac{\theta^r}{r!},\qquad
 \mathbb{E}_{\pi}(X)=\operatorname{Var}_{\pi}[X]=\theta.
 \label{eq:poisson}
\end{equation}
\begin{IEEEproof}
Detailed balance gives $\lambda\pi_r=(r+1)\mu\pi_{r+1}$. Iterating this relation yields
\[
 \pi_r=\pi_0\prod_{j=1}^{r}\frac{\lambda}{j\mu}
      =\pi_0\frac{\theta^r}{r!}.
\]
Normalization then gives
\[
 1=\sum_{r=0}^{\infty}\pi_r
   =\pi_0\sum_{r=0}^{\infty}\frac{\theta^r}{r!}
   =\pi_0e^{\theta},
\]
so $\pi_0=e^{-\theta}$. The process is irreducible. It is nonexplosive because births have a finite constant rate and deaths are bounded by the initial population plus cumulative births. The normalizable invariant distribution therefore establishes positive recurrence and a unique stationary law. Its mean and variance are both $\theta$.
\end{IEEEproof}

For a deterministic initial value $X(0)=x_0$, surviving initial individuals and surviving immigrants also give the exact transient representation
\begin{equation}
 X(t)\ \overset{d}{=}\ \operatorname{Bin}(x_0,e^{-\mu t})+
 \operatorname{Pois}\!\left(\theta[1-e^{-\mu t}]\right),
 \label{eq:transient}
\end{equation}
where $\operatorname{Bin}(b,p)$ denotes a binomial variable with $b$ trials and success probability $p$, and $\operatorname{Pois}(v)$ denotes a Poisson variable with mean $v$. The binomial term counts the $x_0$ initial individuals that survive until $t$, each independently with probability $e^{-\mu t}$. The Poisson term counts individuals born during $(0,t]$ that remain alive at $t$. A birth at time $u$ survives with probability $e^{-\mu(t-u)}$; thinning the Poisson arrivals therefore gives mean $\int_0^t\lambda e^{-\mu(t-u)}\,du=\theta(1-e^{-\mu t})$. The two counts are independent because the arrival process and all individual lifetimes are independent. Consequently,
\begin{align}
 \mathbb{E}[X(t)]&=x_0e^{-\mu t}+\theta(1-e^{-\mu t}),\nonumber\\
 \operatorname{Var}[X(t)]&=x_0e^{-\mu t}(1-e^{-\mu t})+\theta(1-e^{-\mu t}).
 \label{eq:moments}
\end{align}
The relaxation time is $1/\mu$. In particular, increasing $\lambda$ at fixed $\mu$ increases the equilibrium mean without changing this exponential relaxation rate.

Applying \eqref{eq:poisson} to the two populations gives $\mathbb{E}[N]=\lambda_n/\mu_n$ and $\mathbb{E}[M]=\lambda_e/\mu_e$. Their clocks are independent, so the stationary population law is the product of these Poisson laws. These results hold for direct deletion and for the inheritance variants considered here. The arrival and lifetime processes determine population stability. Analysis of topological stationarity requires an augmented state containing the connection structure and the histories used for inheritance.

\subsection{Connection Inheritance and Degree Evolution Under IBAD}
\label{sec:retention}

All probabilities below are conditional on the current connection structure and histories, immediately before a deletion. For a departing node $i$ and a connected hyperedge $e$, the probability of successfully establishing a replacement connection is
\begin{equation}
 r^n_{ie}=\sum_{j\ne i}q^n_{ij}(1-B_{je}).
 \label{eq:success}
\end{equation}
Deleting node $i$ removes its $k_i$ connections. Inheritance establishes an expected $K_i^n=\sum_e B_{ie}r^n_{ie}$ replacement connections, giving an expected net loss of $k_i-K_i^n$ connections. The loss accounts for both rejected duplicate proposals and the absence of eligible candidates.

Let $c_{ij}=\sum_e B_{ie}B_{je}$ be the number of hyperedges shared by nodes $i$ and $j$. When $i$ is deleted, $k_i-c_{ij}$ of its connected hyperedges can establish a new connection with $j$. Independent draws from the fixed proposal distribution select $j$ with probability $q^n_{ij}$ for each of these hyperedges. Hence, the number $G_{ij}$ of new connections received by $j$ satisfies
\begin{align}
 G_{ij}&\sim\operatorname{Bin}(k_i-c_{ij},q^n_{ij}),\nonumber\\
 \mathbb{E}[G_{ij}]&=q^n_{ij}(k_i-c_{ij}),\nonumber\\
 \Pr(G_{ij}\ge1)&=1-(1-q^n_{ij})^{k_i-c_{ij}}.
 \label{eq:recipient}
\end{align}

Shared hyperedges reduce the number of connections that $j$ can gain from $k_i$ to $k_i-c_{ij}$. Equation~\eqref{eq:recipient} gives both the expected gain and the probability of receiving at least one connection. For deletion of hyperedge $f$, the dual formulas are
\begin{equation}
 r^e_{if}=\sum_{h\ne f}q^e_{fh}(1-B_{ih}),\qquad K_f^e=\sum_i B_{if}r^e_{if}.
 \label{eq:edge-success}
\end{equation}
Thus, $0\le K_i^n\le k_i$ and $0\le K_f^e\le s_f$. Direct deletion gives $K_i^n=K_f^e=0$. At a fixed pre-deletion state, sampling from feasible candidates establishes a replacement connection whenever this set is nonempty. Under single-proposal inheritance, an existing connection causes rejection.

Let
\begin{equation}
 b(H)=\lambda_n\min(m_n,M)+\lambda_e\min(m_e,N)
 \label{eq:birth-incidences}
\end{equation}
be the rate of connection creation through births. Weighting the expected net loss of connections at each deletion by its death rate yields the exact first-moment balance
\begin{align}
 \frac{d\mathbb{E}(L)}{dt}=\mathbb{E}\bigg(b(H)
 &-\mu_n\sum_i(k_i-K_i^n)\nonumber\\
 &-\mu_e\sum_f(s_f-K_f^e)\bigg).
 \label{eq:L-drift}
\end{align}
This equation expresses the balance between connection creation and net loss of connections. If a stationary topological regime with finite first moments exists, define the connection retention fractions
\begin{equation}
 \eta_n=\frac{\mathbb{E}(\sum_i K_i^n)}{\mathbb{E}(L)},\qquad
 \eta_e=\frac{\mathbb{E}(\sum_f K_f^e)}{\mathbb{E}(L)}.
 \label{eq:eta}
\end{equation}
Here, connection retention measures replacement of removed connections through inheritance. Since $\sum_i k_i=\sum_f s_f=L$, setting $d\mathbb{E}(L)/dt=0$ in \eqref{eq:L-drift} and substituting \eqref{eq:eta} gives
\[
 \mathbb{E}(b(H))
 =\big[\mu_n(1-\eta_n)+\mu_e(1-\eta_e)\big]\mathbb{E}(L).
\]
Solving for $\mathbb{E}(L)$, provided the coefficient on the right is positive, yields
\begin{equation}
 \mathbb{E}(L)=\frac{\mathbb{E}(b(H))}{\mu_n(1-\eta_n)+\mu_e(1-\eta_e)}.
 \label{eq:L-stationary}
\end{equation}
For direct deletion, this reduces to $\mathbb{E}(b(H))/(\mu_n+\mu_e)$. The connection retention fractions in \eqref{eq:eta} are determined jointly by histories and connection overlaps. Since $L\le NM$, stationary independent population counts also imply $\mathbb{E}(L)\le(\lambda_n/\mu_n)(\lambda_e/\mu_e)$. Establishing existence of a stationary law for the augmented topological process requires further analysis.

For completeness, consider the degree of node $i$ while it remains alive. Let $A_i^e$ be its inclusion probability in the distinct-target sample at a hyperedge birth. Its conditional drift is the expected instantaneous rate of degree change from events other than its own removal:
\begin{align}
 g_i^n={}&\lambda_e A_i^e
 +\mu_n\sum_{u\ne i}q^n_{ui}\sum_f B_{uf}(1-B_{if})\nonumber\\
 &-\mu_e\sum_f B_{if}(1-r^e_{if}).
 \label{eq:degree-drift}
\end{align}
The corresponding drift for a surviving hyperedge $e$ is
\begin{align}
 g_e^e={}&\lambda_n A_e^n
 +\mu_e\sum_{f\ne e}q^e_{fe}\sum_i B_{if}(1-B_{ie})\nonumber\\
 &-\mu_n\sum_i B_{ie}(1-r^n_{ie}),
 \label{eq:size-drift}
\end{align}
where $A_e^n$ is its inclusion probability at node birth. Both inclusion probabilities refer to selection at least once during the full sequence of draws without replacement. In each expression, the first term describes connections gained through preferential attachment at opposite-type births. The second describes connections gained by inheriting from departing same-type individuals; the factors $1-B$ restrict gains to connections that are absent. The third describes connections lost at opposite-type deaths, weighted by the probability $1-r$ that inheritance fails to establish a replacement connection.

One same-type death can produce several connection gains to a survivor, requiring multiple-jump terms in a degree master equation. Deriving a stationary degree distribution or its tail exponent requires a closure for the joint history and overlap statistics. The present analysis establishes conditional connection retention and moment relations; distribution shapes are assessed numerically.

\section{Simulations}
\label{sec:simulations}

All simulations are implemented in Python. Unless stated otherwise, the topology experiments use single-proposal IBAD with raw-history similarity, $a_n=a_e=1$, $h_{\min}=100$, and $\sigma=1$. Each run begins with a burn-in period of $10^5$ events before observations are collected. The network then evolves for another $2\times10^4$ events, during which its topological properties are sampled every 400 events, yielding 50 samples. Unless otherwise specified in the experiment-specific captions, results reported in the remainder of this section are obtained by averaging all samples collected from ten independent runs for each parameter setting. Node-degree and hyperedge-size distributions are obtained by averaging the corresponding counts over the observation window and across independent runs, followed by normalization.

\subsection{Stationarity of Node and Hyperedge Populations}

\begin{figure}[!ht]
\centering
\includegraphics[scale=0.16]{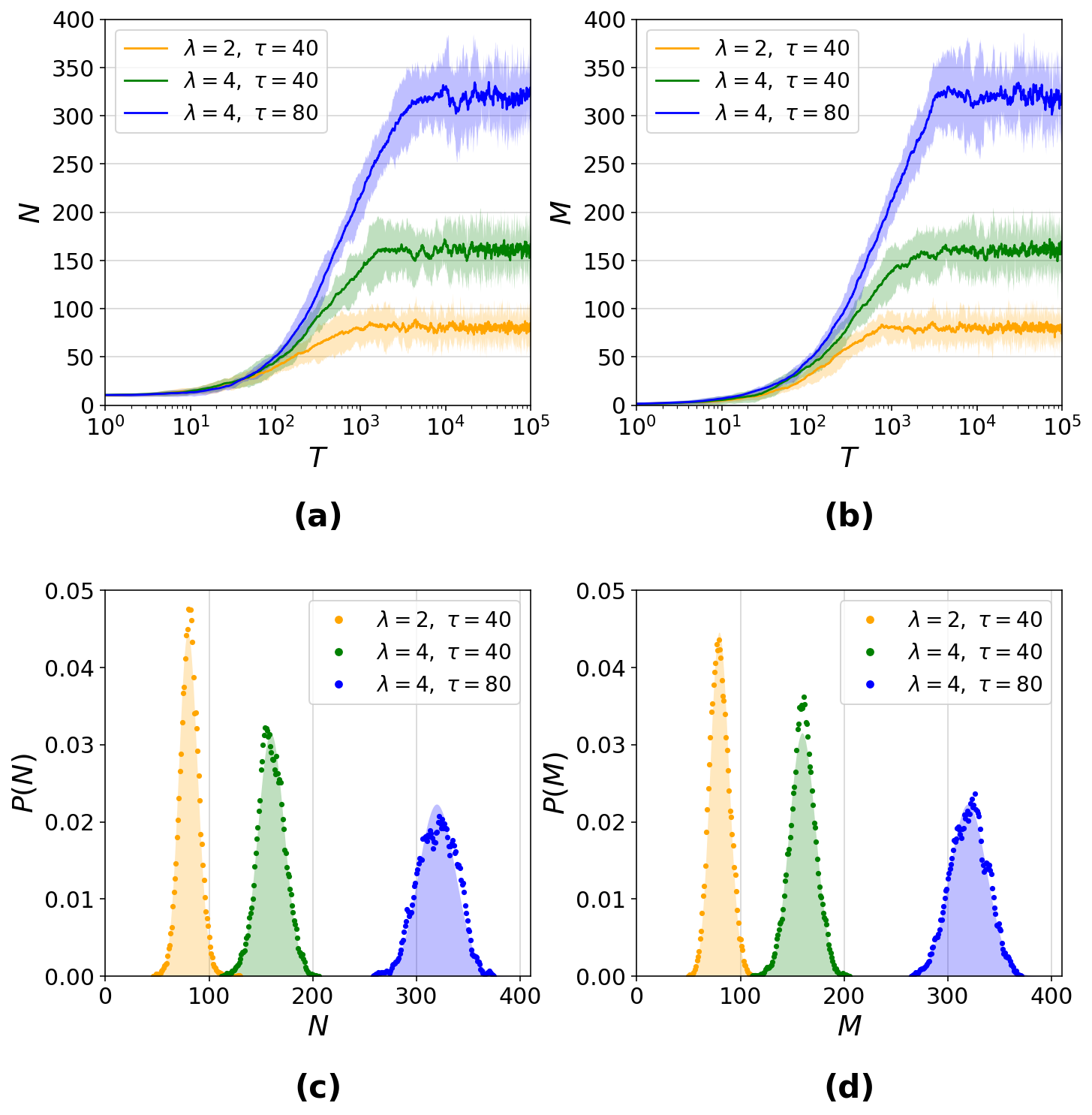}
\caption{\label{fig2}
\textbf{Temporal evolution and stationary distributions of node and hyperedge populations.} Panels (a) and (b) show node and hyperedge populations against the number of completed birth or death events $T$. The node and hyperedge birth rates are equal to $\lambda$, and their mean lifetimes are equal to $\tau=1/\mu$. The orange, green, and blue curves correspond to $(\lambda,\tau)=(2,40),(4,40),(4,80)$, respectively. Solid curves show the mean across independent runs at the same event index; shaded bands indicate $\pm2$ sample standard deviations across runs. In (c) and (d), dots show simulated population distributions. Shaded profiles show the theoretical Poisson probabilities with mean $\lambda\tau$.}
\end{figure}

Figure~\ref{fig2} shows the temporal evolution and stationary distributions of node and hyperedge populations, with $N(0)=10$ and $M(0)=1$. The parameter combinations are $(\lambda,\tau)=(2,40),(4,40),(4,80)$, with equal node and hyperedge parameters and $\tau=1/\mu$.

For the population distributions in Fig.~\ref{fig2}(c,d), states are weighted by their physical holding times during the observation window. The resulting distribution is normalized within each run and then averaged across independent runs. 

Both populations increase initially and then fluctuate around their theoretical stationary means. Figure~\ref{fig2}(c) and Fig.~\ref{fig2}(d) further indicate that smaller networks exhibit smaller fluctuations, whereas larger networks show larger variance and larger absolute fluctuations.
 
Table~\ref{tab:table1} shows that simulated population means agree with the birth-death predictions, with relative errors below 0.2\% in all cases.

\begin{table}[!t]
\centering
\caption{Simulated Population Means and Theoretical Expectations}\label{tab:table1}
\small\setlength{\tabcolsep}{3.5pt}\renewcommand{\arraystretch}{1.12}
\begin{tabular}{@{}llrrrr@{}}
\toprule
$(\lambda,\tau)$ & Population & Theory & \shortstack{Simulation\\mean} & \shortstack{Absolute\\error} & \shortstack{Relative\\error}\\
\midrule
\multirow{2}{*}{(2,40)} & Node & 80 & 79.978 & 0.022 & 0.028\%\\
 & Hyperedge & 80 & 80.089 & 0.089 & 0.112\%\\
\addlinespace[3pt]
\multirow{2}{*}{(4,40)} & Node & 160 & 159.841 & 0.159 & 0.100\%\\
 & Hyperedge & 160 & 159.932 & 0.068 & 0.042\%\\
\addlinespace[3pt]
\multirow{2}{*}{(4,80)} & Node & 320 & 319.947 & 0.053 & 0.017\%\\
 & Hyperedge & 320 & 320.025 & 0.025 & 0.008\%\\
\bottomrule
\end{tabular}
\end{table}

\subsection{Topological Characteristics of Networks With Inheritance-Based Deletion}

\begin{figure}[!t]
\centering
\subfloat[Node degree; varying $\lambda$]{\includegraphics[width=.47\columnwidth]{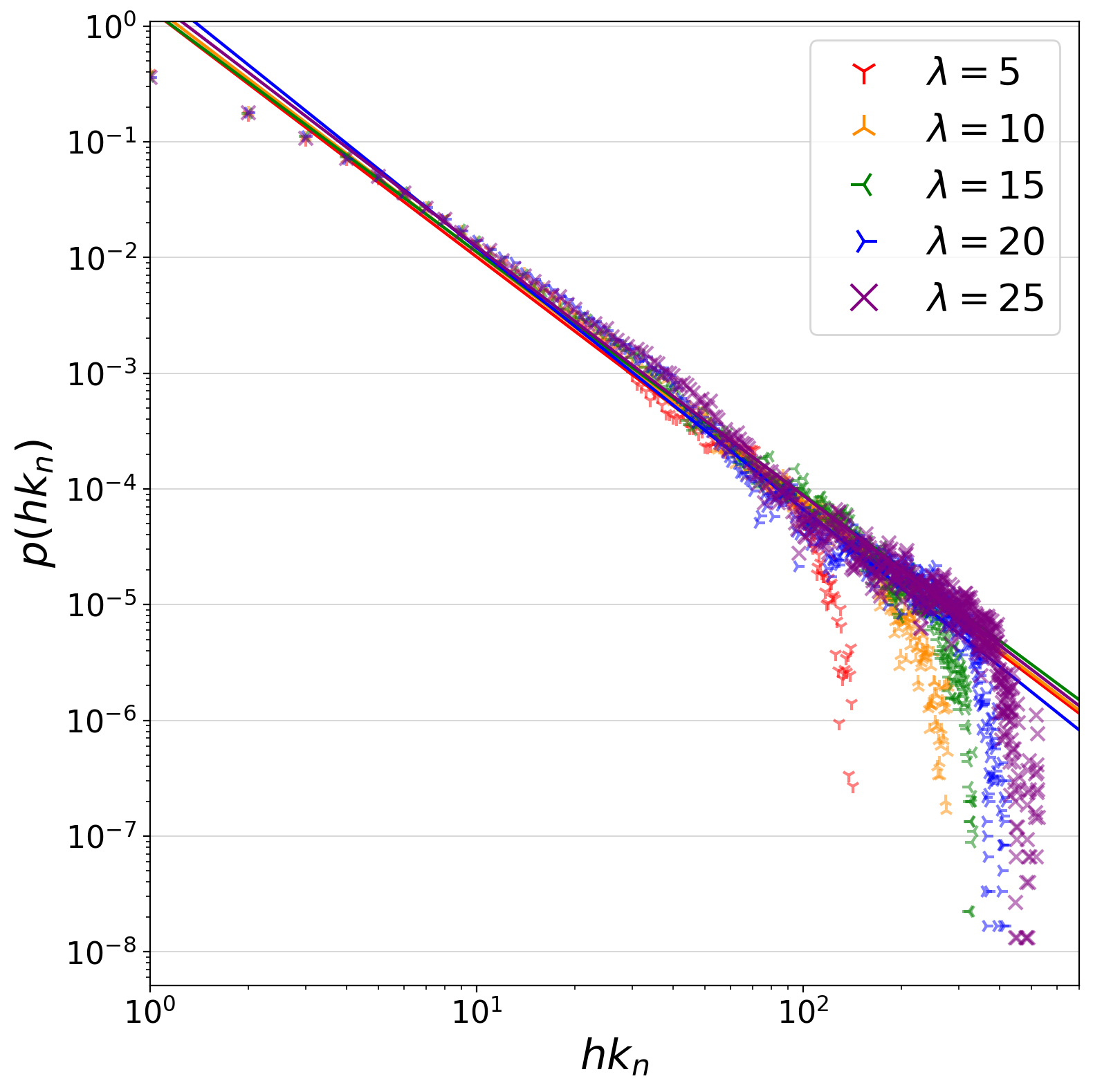}}\hfill
\subfloat[Hyperedge size; varying $\lambda$]{\includegraphics[width=.47\columnwidth]{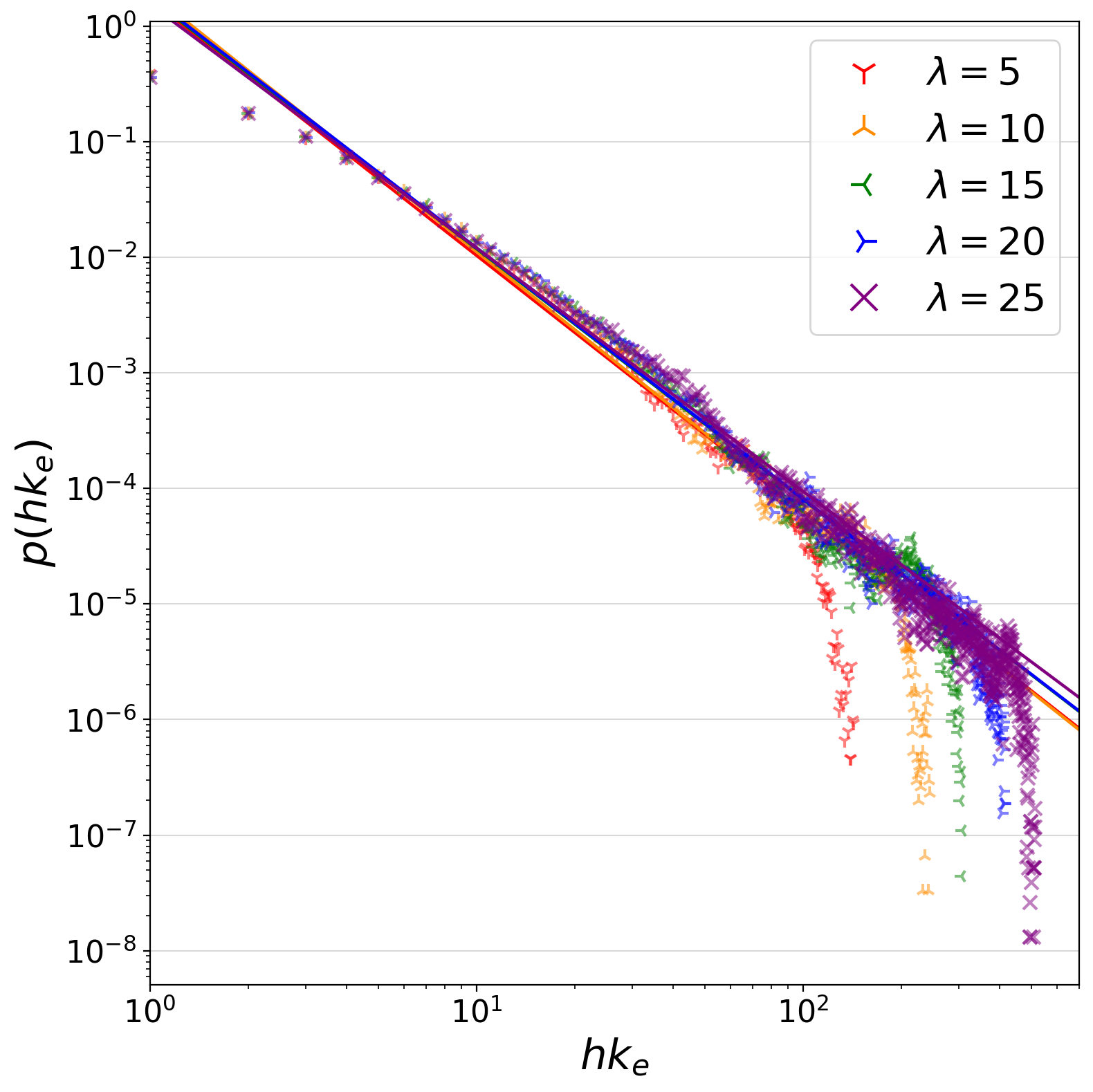}}\\[1pt]
\subfloat[Node degree; varying $\tau$]{\includegraphics[width=.47\columnwidth]{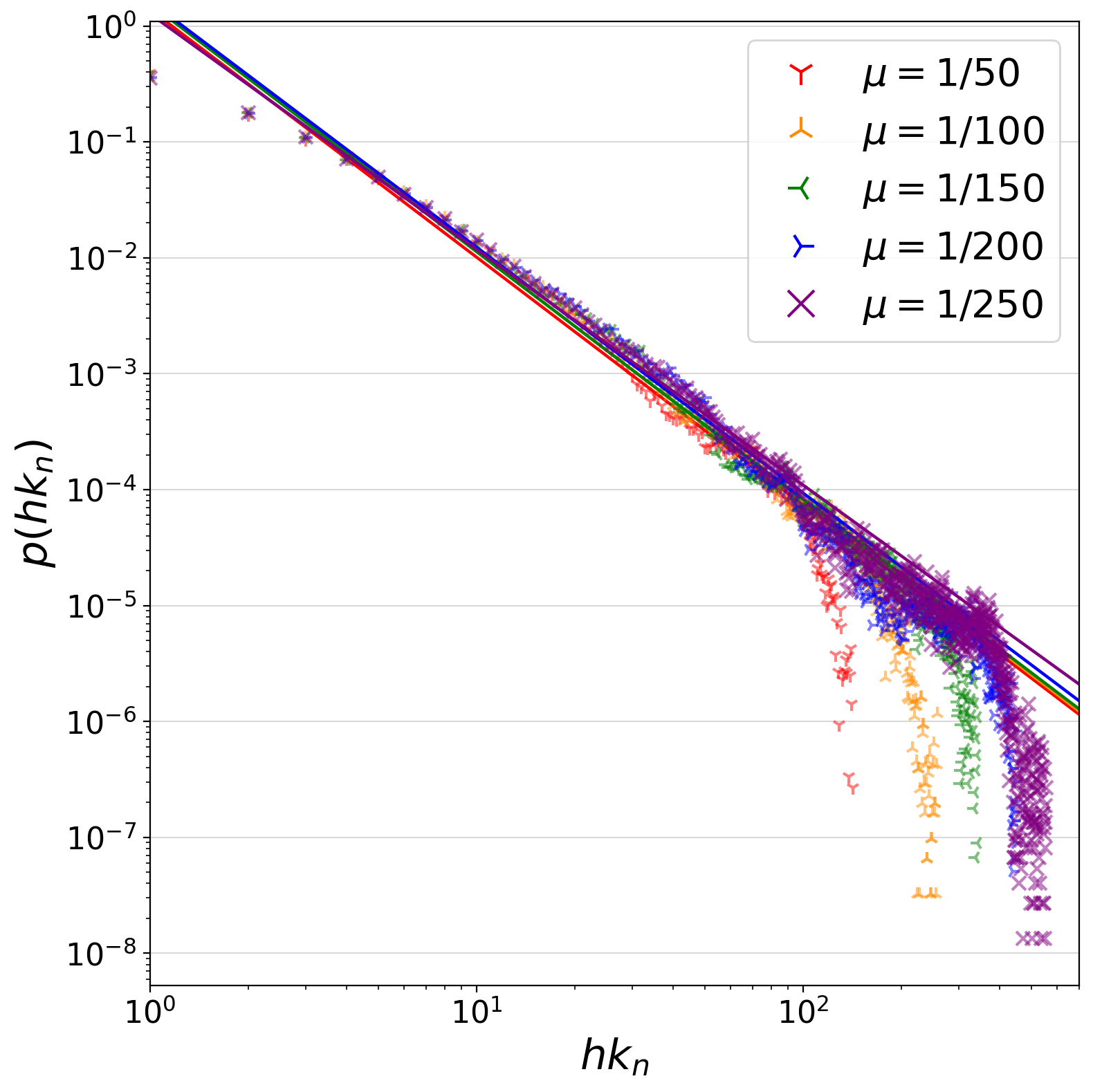}}\hfill
\subfloat[Hyperedge size; varying $\tau$]{\includegraphics[width=.47\columnwidth]{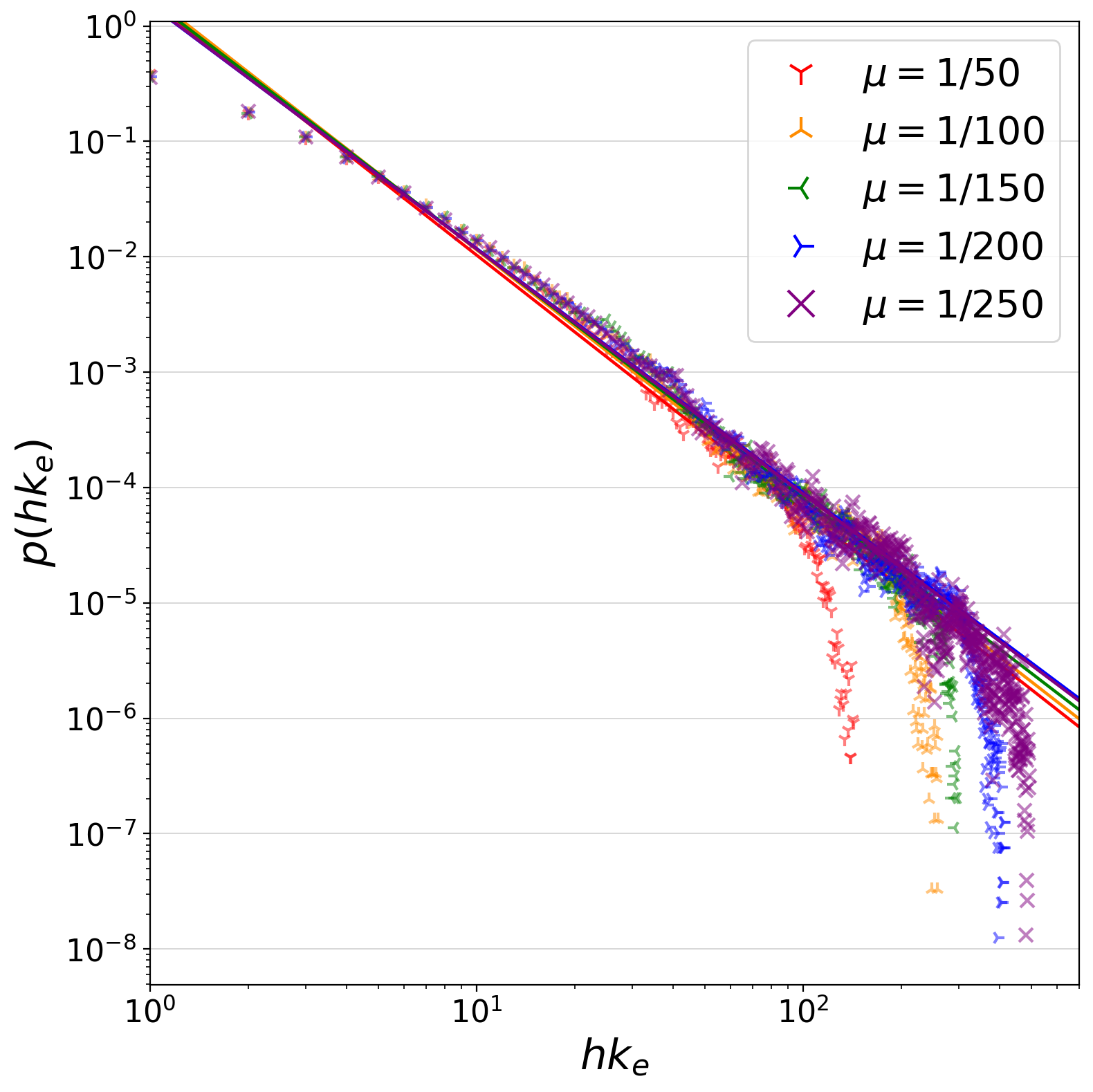}}\\[1pt]
\caption{\textbf{Effects of birth rates and mean lifetimes on node-degree and hyperedge-size distributions.} In (a,b), $\tau_n=\tau_e=50$; in (c,d), $\lambda_n=\lambda_e=5$ and the plotted $\mu$ is $1/\tau$. All panels use $m_n=m_e=1$. Slopes in Table~\ref{tab:table2} use $1\le k\le100$; line segments outside that interval are extrapolations.}\label{fig3}
\end{figure}

\textit{Birth rates and mean lifetimes:} Figure~\ref{fig3} compares degree distributions for $\lambda_n=\lambda_e=\lambda\in\{5,10,15,20,25\}$ at $\tau_n=\tau_e=50$, and for $\tau_n=\tau_e=\tau\in\{50,100,150,200,250\}$ at $\lambda_n=\lambda_e=5$. All simulations use $m_n=m_e=1$. The distributions show approximately linear decay over intermediate ranges on logarithmic axes, followed by a sharper decrease in the far tail. The intermediate-range slopes vary comparatively little across the tested settings, while differences are more apparent in the extent and shape of the tail.

The sharp downturn in the far tail may reflect both finite-size constraints and rejected inheritance proposals. At any instant, node degrees and hyperedge sizes are bounded by the numbers of existing hyperedges and nodes, respectively. Under the symmetric settings considered here, both stationary mean populations equal $\lambda\tau$, so increasing $\lambda$ or $\tau$ can extend the range of attainable degrees. Single-proposal inheritance can further limit connection gains when degree-history similarity favors highly connected recipients whose existing connections overlap with those offered for inheritance. Rejection of these duplicate proposals may therefore contribute to the tail downturn. Its magnitude depends on the proposal probabilities and the actual connection overlap.

To compare the decay of the degree distributions, we perform
a least-squares fit in logarithmic coordinates:
\begin{equation}
	\log\widehat p(k)=c-\gamma\log k,
	\label{eq:fit}
\end{equation}
where $c$ is the intercept and $-\gamma$ is the fitted slope.
Only positive-probability bins within the specified degree
interval are included. Table~\ref{tab:table2} quantifies this observation. The fitted slopes show comparatively weak sensitivity to birth and death rates over the tested range.

\begin{table}[!t]
\centering
\caption{Descriptive Decay Parameters for Birth-Rate and Lifetime Sweeps}\label{tab:table2}
\small
\setlength{\tabcolsep}{6.5pt}
\renewcommand{\arraystretch}{1.12}
\begin{tabular}{@{}ccc@{\hspace{15pt}}ccc@{}}
\toprule
\multirow{2}{*}{$\lambda$}
& \multicolumn{2}{c@{\hspace{15pt}}}{$\tau=50$}
& \multirow{2}{*}{$\tau$}
& \multicolumn{2}{c@{}}{$\lambda=5$}\\
\cmidrule(lr){2-3}\cmidrule(l){5-6}
\phantom{$(\tau=50)$}
& $\gamma_n$ & $\gamma_e$
& \phantom{$(\lambda=5)$}
& $\gamma_n$ & $\gamma_e$\\
\midrule
5 & 2.138 & 2.217 & 50 & 2.138 & 2.217\\
10 & 2.145 & 2.242 & 100 & 2.152 & 2.204\\
15 & 2.098 & 2.170 & 150 & 2.139 & 2.165\\
20 & 2.260 & 2.172 & 200 & 2.122 & 2.114\\
25 & 2.152 & 2.108 & 250 & 2.033 & 2.123\\
\bottomrule
\end{tabular}
\par\smallskip\begin{minipage}{0.97\columnwidth}\footnotesize $m_n=m_e=1$; fit range $1\le k\le100$. The decay parameters are descriptive ordinary least-squares summaries of averaged distributions.\end{minipage}
\end{table}

\begin{figure*}[!t]
\centering
\subfloat[$m_e=1$; node]{\includegraphics[width=.235\textwidth]{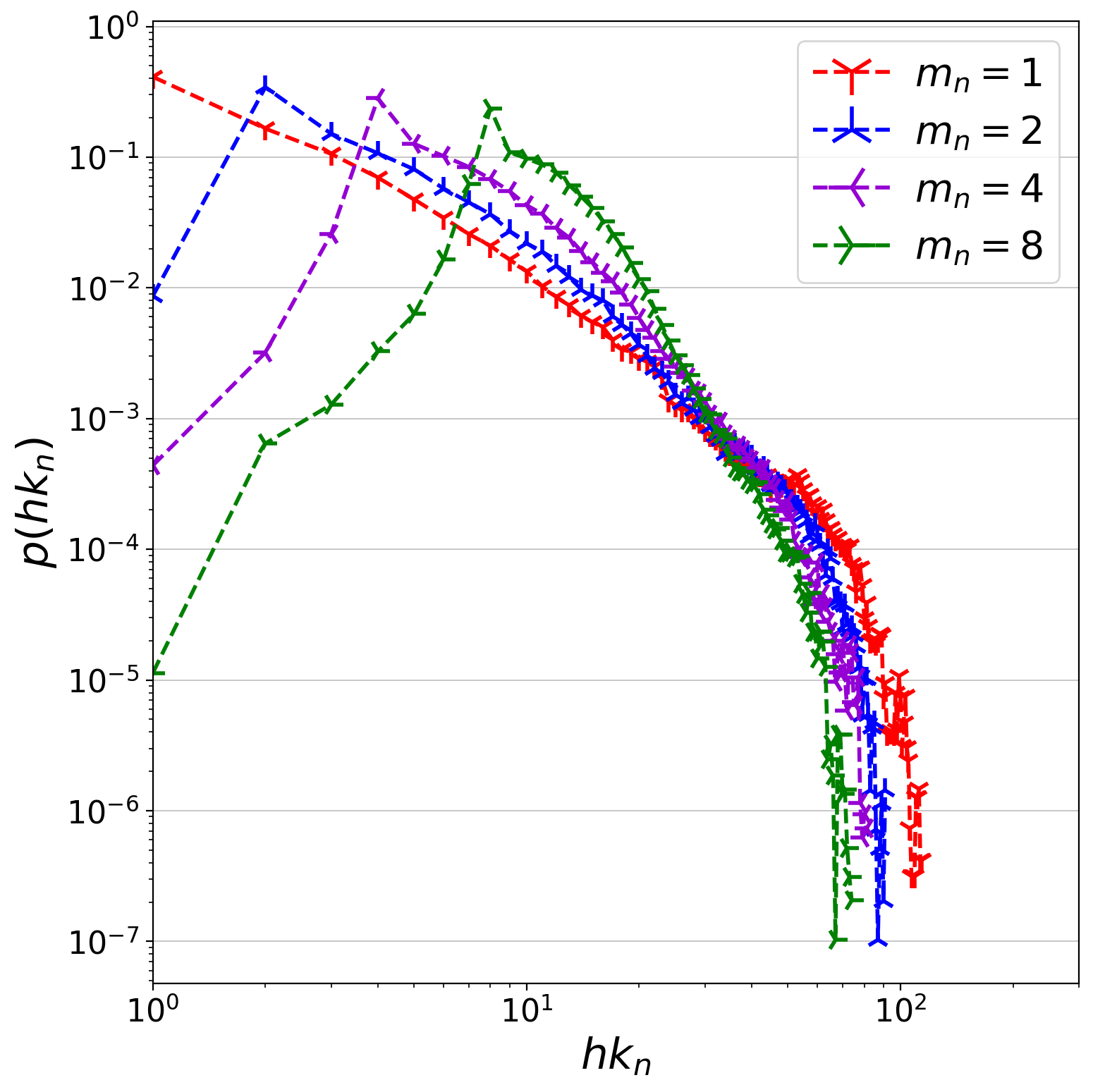}}\hfill
\subfloat[$m_e=2$; node]{\includegraphics[width=.235\textwidth]{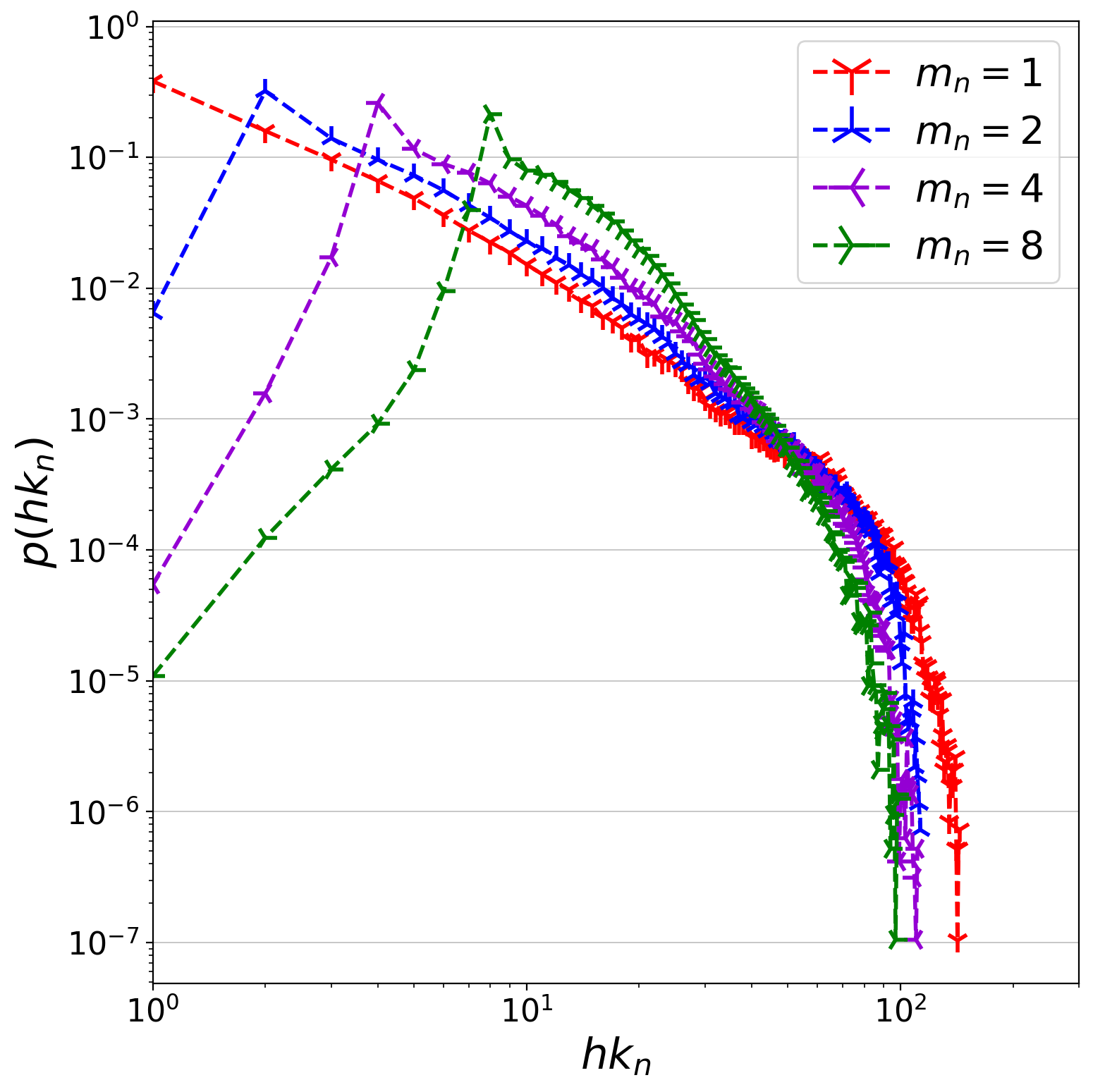}}\hfill
\subfloat[$m_e=4$; node]{\includegraphics[width=.235\textwidth]{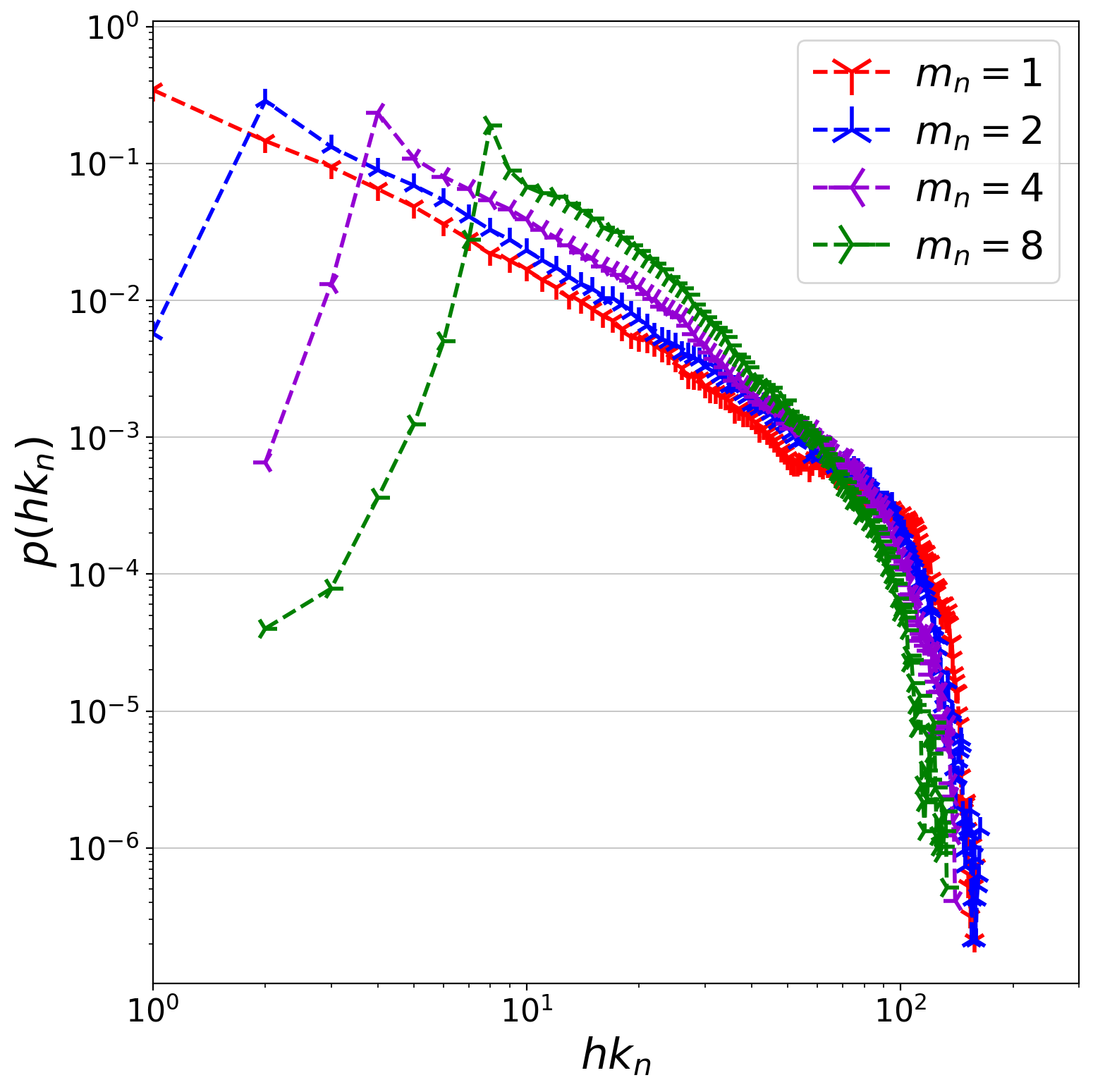}}\hfill
\subfloat[$m_e=8$; node]{\includegraphics[width=.235\textwidth]{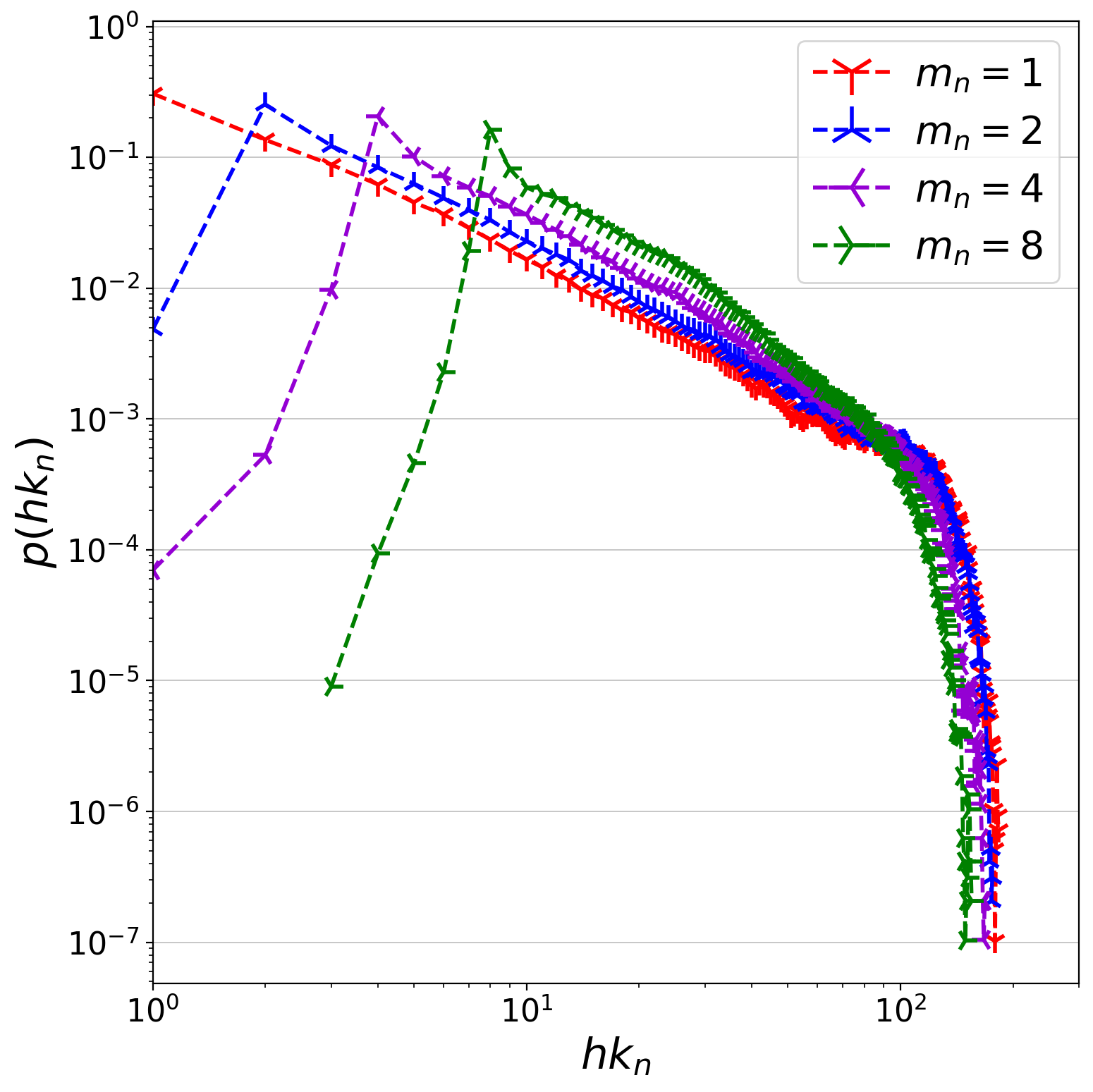}}\\[1pt]
\subfloat[$m_e=1$; hyperedge]{\includegraphics[width=.235\textwidth]{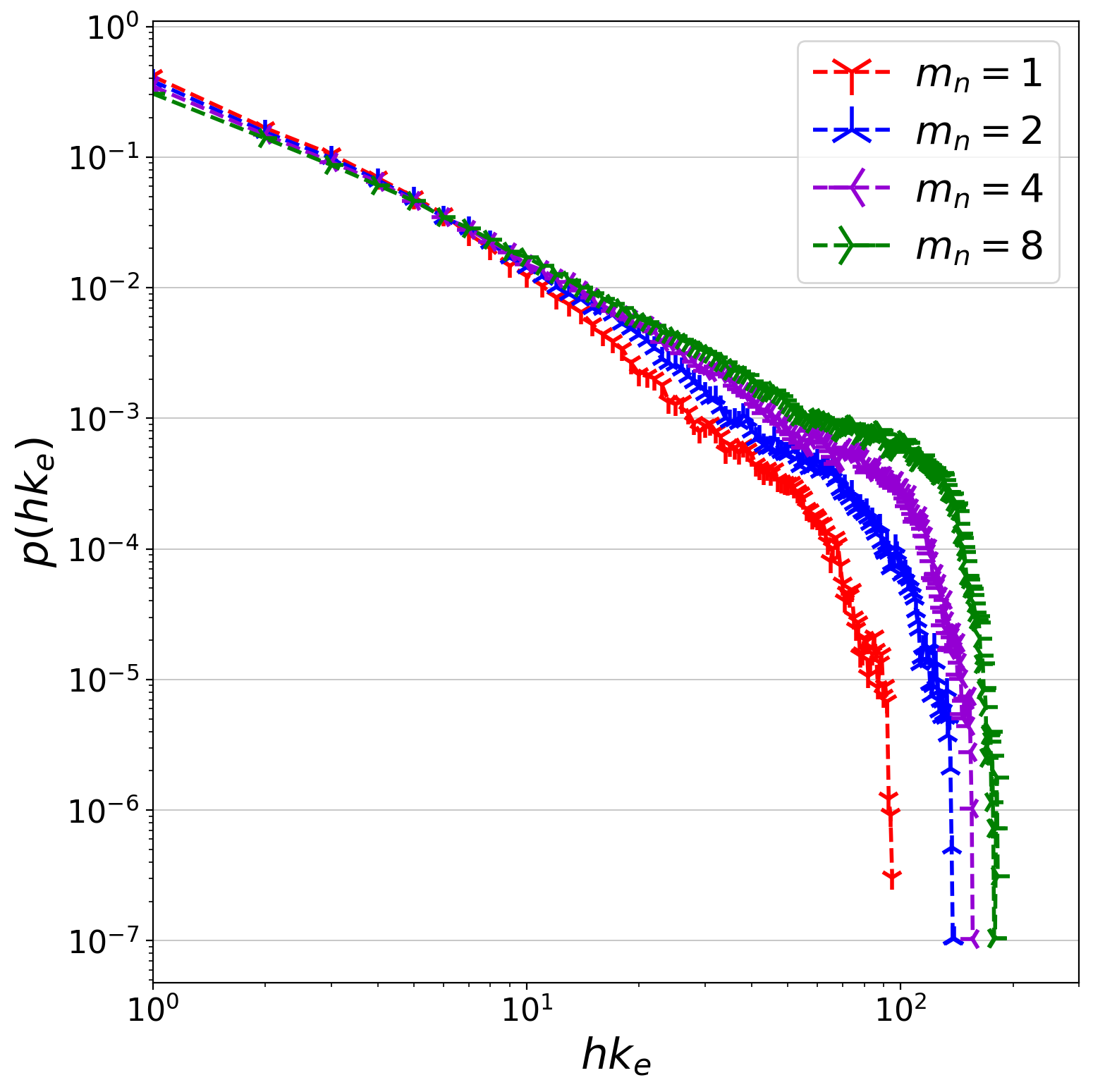}}\hfill
\subfloat[$m_e=2$; hyperedge]{\includegraphics[width=.235\textwidth]{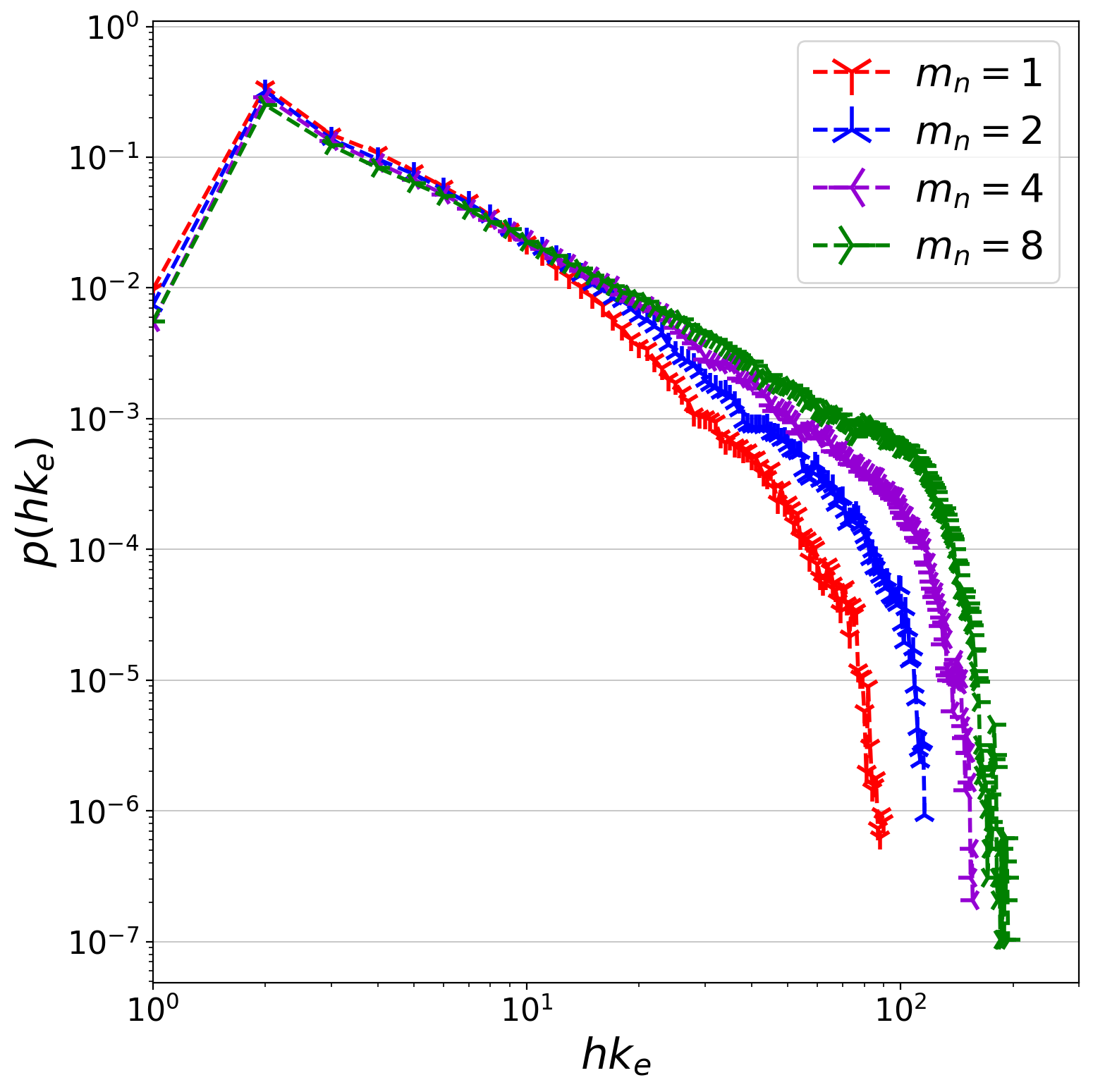}}\hfill
\subfloat[$m_e=4$; hyperedge]{\includegraphics[width=.235\textwidth]{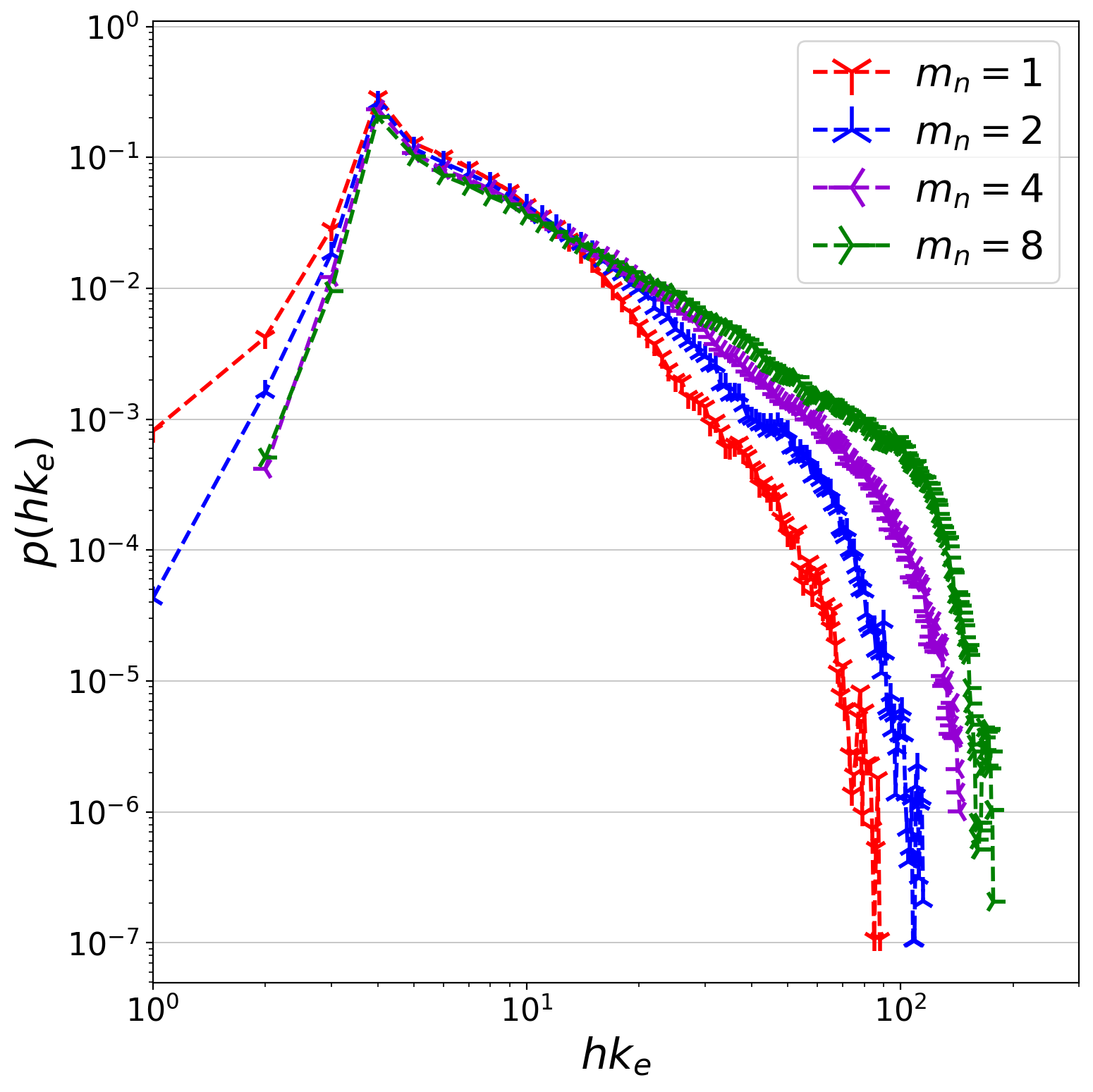}}\hfill
\subfloat[$m_e=8$; hyperedge]{\includegraphics[width=.235\textwidth]{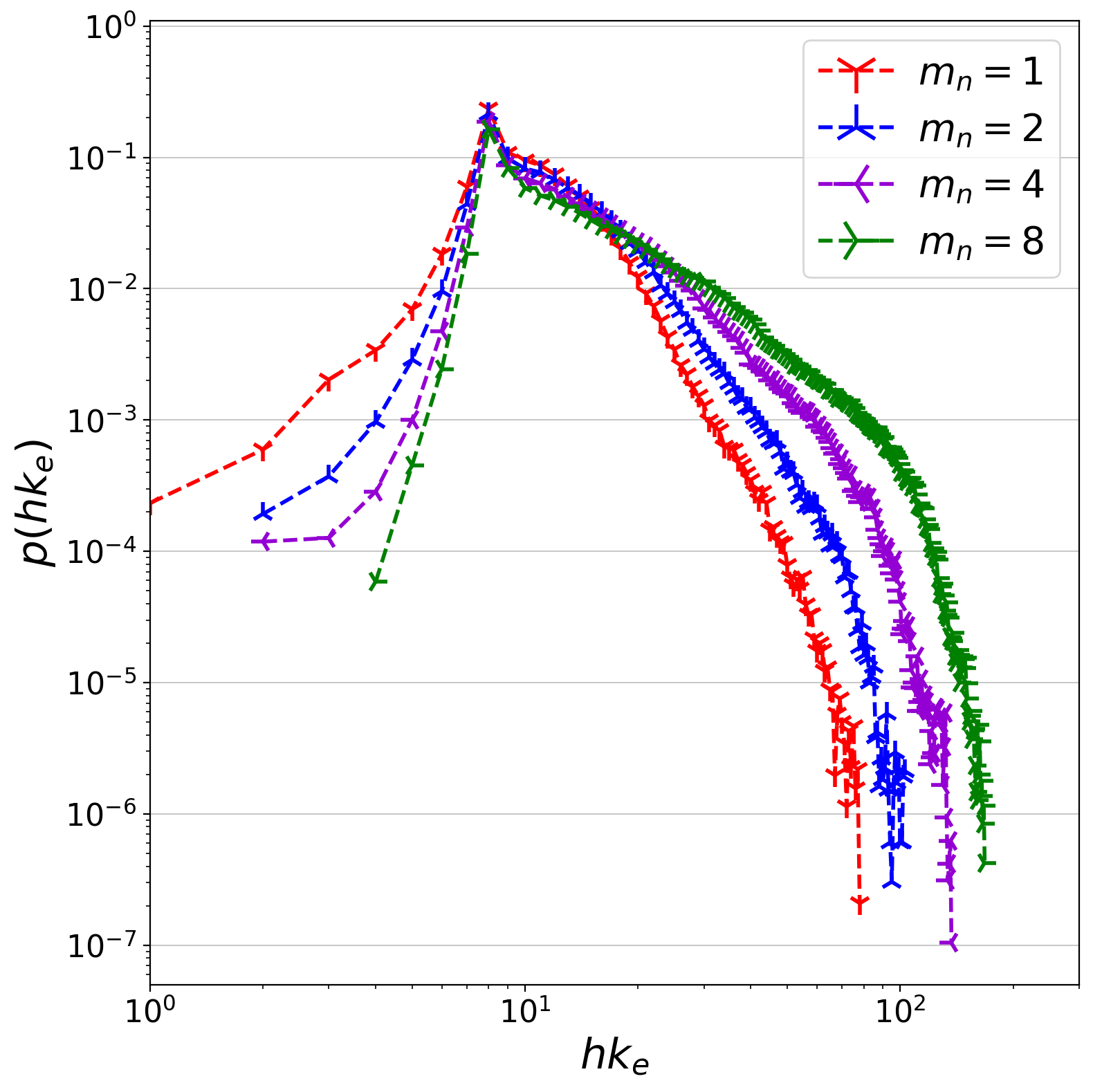}}\\[1pt]
\caption{\textbf{Effects of attachment parameters on node-degree and hyperedge-size distributions.} Panels (a)-(d) show node-degree distributions and (e)-(h) hyperedge-size distributions at fixed $m_e=1,2,4,8$, respectively. Each panel compares $m_n=1,2,4,8$. The other settings are $\lambda_n=\lambda_e=2$ and $\tau_n=\tau_e=80$, with three runs per pair.}\label{fig5}
\end{figure*}

\textit{Attachment parameters:} Figure~\ref{fig5} fixes $m_e$ within each pair of panels and varies $m_n\in\{1,2,4,8\}$. The settings are $\lambda_n=\lambda_e=2$ and $\tau_n=\tau_e=80$. Increasing $m_n$ directly adds connections at node birth; increasing $m_e$ directly changes newborn hyperedge size. Their subsequent effects are coupled through opposite-type deaths, inheritance, and duplicate rejection, as shown by \eqref{eq:degree-drift} and \eqref{eq:size-drift}. The results show that $P(hk_n)$ exhibits approximately power-law decay over a certain degree range when $m_n=1$ or $2$. Similarly, $P(hk_e)$ exhibits approximately power-law decay over a certain degree range when $m_e=1$ or $2$. For the node-degree distributions, a more pronounced low-degree peak appears as $m_n$ increases to 4 or 8. For the hyperedge-size distributions, increasing $m_e$ shifts the peak toward larger sizes. Beyond these peaks, the distributions show approximately power-law decay over part of the decreasing range. This shape resembles the subnormal distributions studied by Feng et al.~\cite{feng2017} in evolving networks with variable elements. In the present model, attachment at birth sets the initial connectivity, while random node and hyperedge lifetimes and inheritance introduce fluctuations in subsequent degree evolution. Together with preferential attachment, these mechanisms may contribute to the emergence of a low-degree peak followed by a broad decreasing region. At fixed $m_e$, the hyperedge-size distributions peak at similar sizes across different values of $m_n$. The node-degree peak shifts toward larger degrees as $m_n$ increases, and its height generally decreases.

\textit{Clustering:} We construct the simple node projection, connecting two nodes when they share at least one hyperedge. Let $T_\triangle$ be its number of distinct triangles, $W=\sum_i\binom{d_i^{\rm proj}}{2}$ its number of centered wedges, and $I$ the number of distinct node triples contained in at least one individual hyperedge. The clustering coefficient is
\begin{equation}
 C=\frac{3(T_\triangle-I)}{W-3I},
 \label{eq:clustering}
\end{equation}
with $C=0$ if the denominator is zero. This removes triples internal to a single group from the projected transitivity calculation, in the spirit of distinguishing group-internal closure from higher-order connectivity \cite{Estrada2006}. Each triple contributes once to $I$, including triples shared by several hyperedges. Connectivity is measured by $R_{\rm LCC}=N_{\rm LCC}/N$ on this same projection, including isolated nodes, and is set to zero for an empty population.

\begin{figure*}[!t]
\centering
\subfloat[$m_n=m_e=1$]{\includegraphics[width=.30\textwidth]{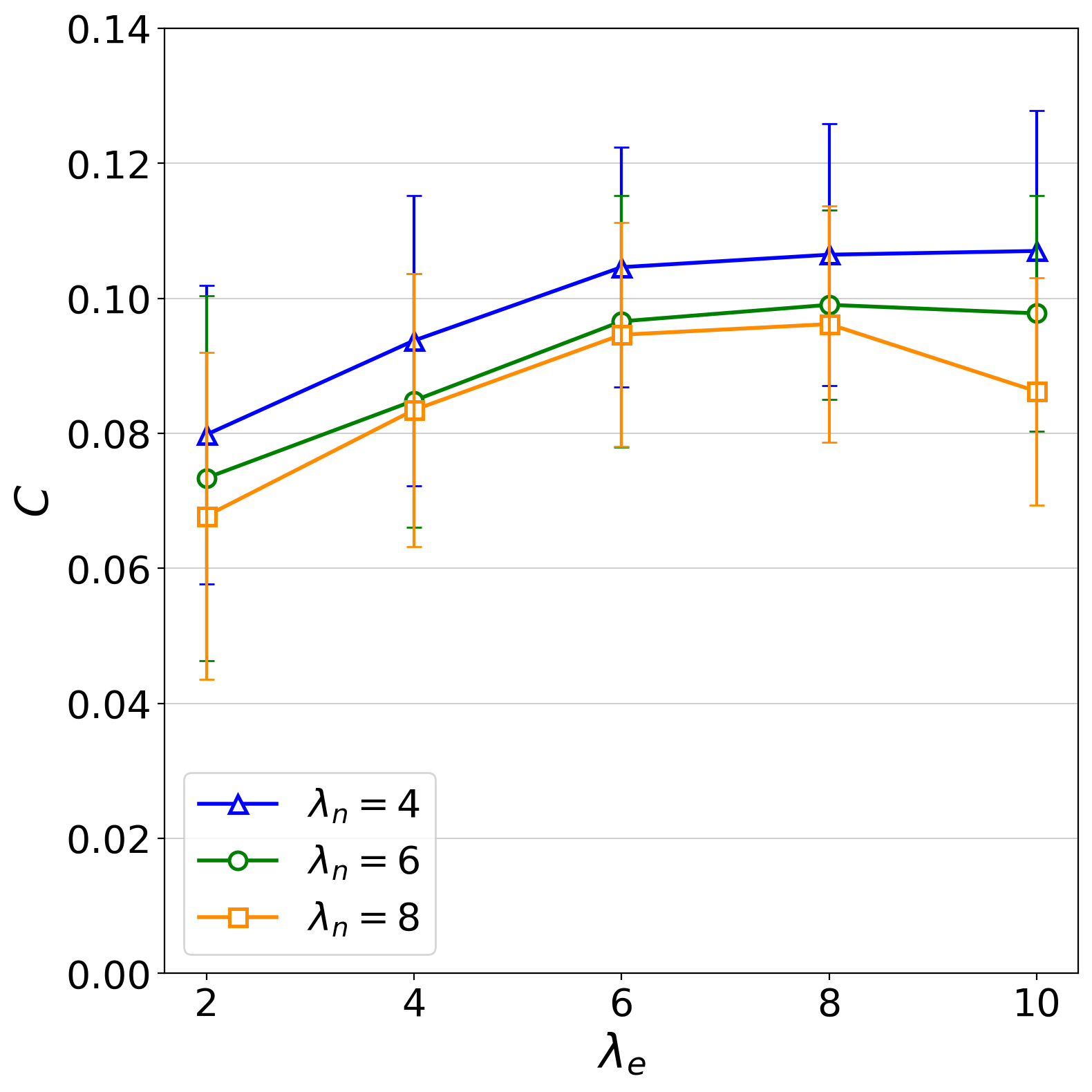}}\hfill
\subfloat[$m_n=m_e=2$]{\includegraphics[width=.30\textwidth]{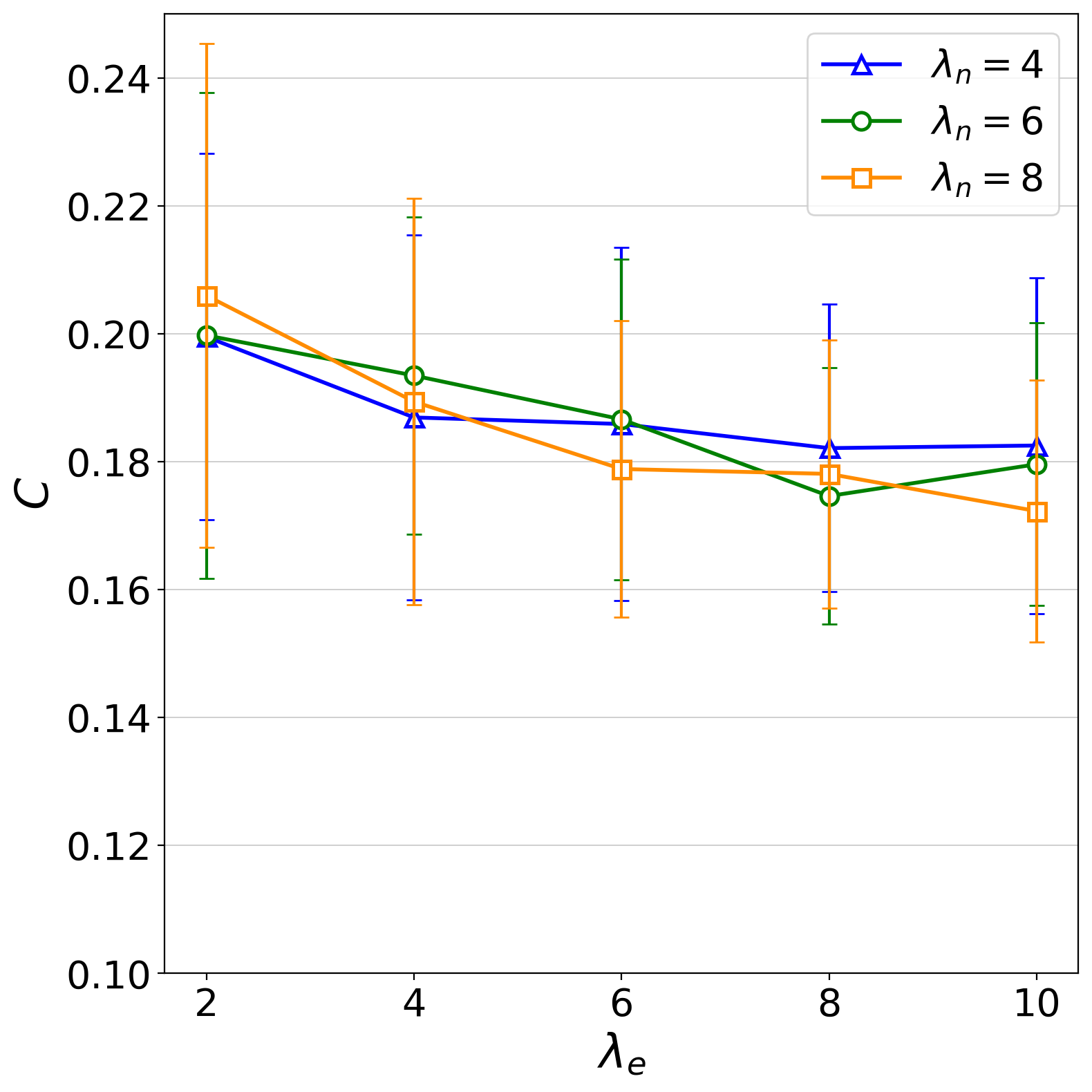}}\hfill
\subfloat[Attachment parameters]{\includegraphics[width=.30\textwidth]{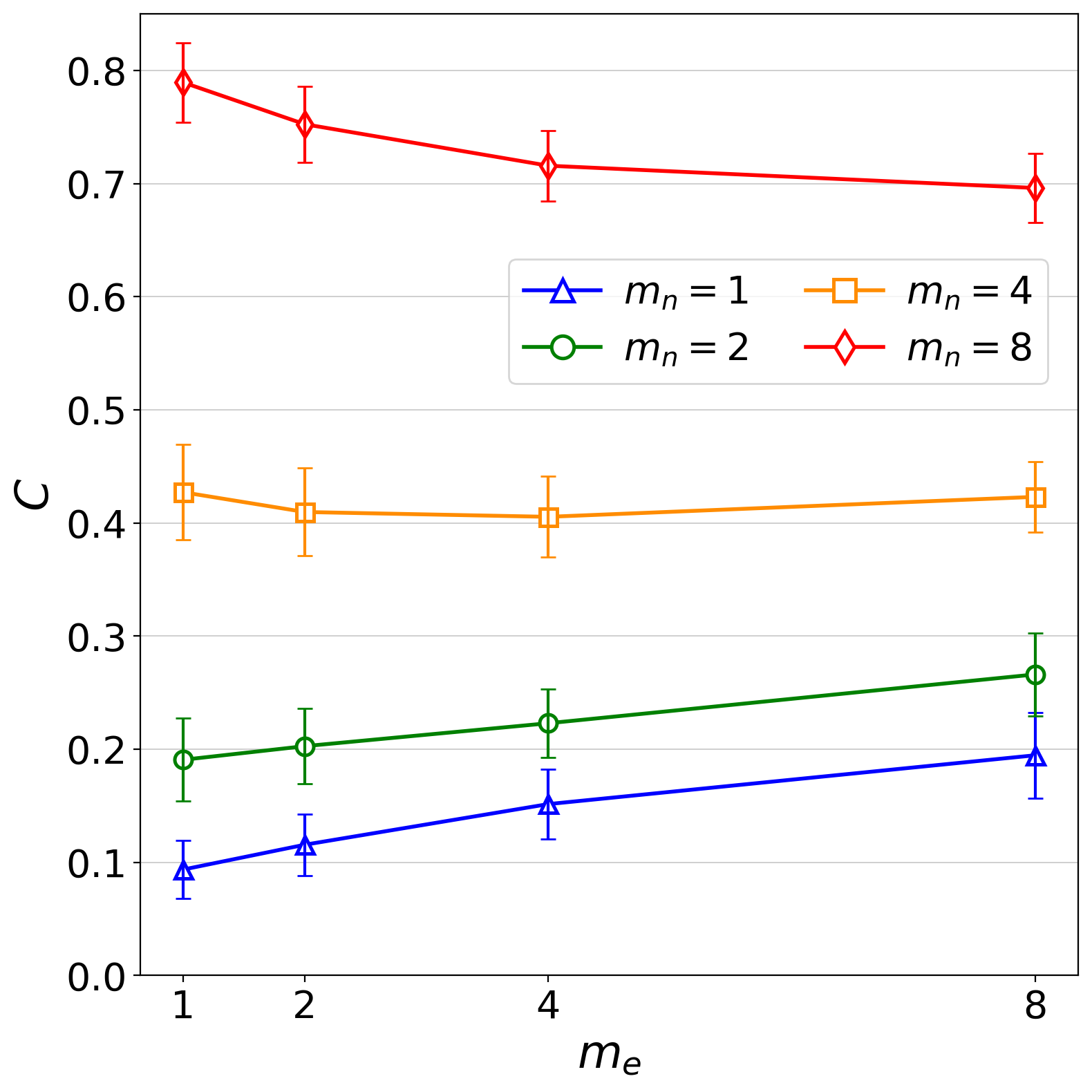}}\\[1pt]
\caption{\textbf{Clustering under changes in birth and attachment parameters.} Panels (a,b) vary $\lambda_e$ and distinguish $\lambda_n$ by separate curves. Panel (c) varies $m_e$ and distinguishes $m_n$. All panels use $\tau_n=\tau_e=80$, i.e., $\mu_n=\mu_e=1/80$. Panels (a) and (b) fix $m_n=m_e=1$ and 2, respectively. Panel (c) fixes $\lambda_n=\lambda_e=2$. Error bars indicate $\pm1$ sample standard deviation.}\label{fig4}
\end{figure*}

Figure~\ref{fig4}(a,b) shows how the birth rates affect the clustering coefficient under two attachment settings. For $m_n=m_e=1$, $C$ generally increases with $\lambda_e$, with smaller changes at higher birth rates. For $m_n=m_e=2$, the mean values show a modest overall decrease. The differences between the $\lambda_n$ curves are small relative to the displayed variability, particularly in Fig.~\ref{fig4}(b).

At fixed mean lifetimes, the birth rates determine the stationary mean populations, $\mathbb{E}(N)=\lambda_n\tau_n$ and $\mathbb{E}(M)=\lambda_e\tau_e$. Changing these rates alters relative population sizes and attachment frequencies, affecting both closed configurations and open two-step paths. Since \eqref{eq:clustering} excludes triples within a single hyperedge, its response depends on connections across hyperedges. The contrasting trends in Fig.~\ref{fig4}(a,b) show that the response of clustering to the hyperedge birth rate depends on the attachment setting.

Figure~\ref{fig4}(c) further shows that increasing $m_n$ raises $C$ at each tested value of $m_e$. A larger $m_n$ allows a newborn node to join more existing hyperedges, potentially promoting closure across hyperedges. Increasing $m_e$ instead places more nodes in each newborn hyperedge. Because within-hyperedge triples are excluded, its effect on $C$ depends on the selected nodes' existing memberships. In the tested settings, $C$ increases with $m_e$ for $m_n=1$ and 2, varies relatively little for $m_n=4$, and decreases for $m_n=8$.

\subsection{Comparison of Deletion Mechanisms and Similarity Measures}
\label{sec:comparison}

We compare direct deletion, uniform single-proposal inheritance, and raw-history IBAD. The uniform rule samples uniformly from all history-eligible same-type survivors. We then retain single-proposal inheritance and compare raw distance with log-transformed distance, shifted Pearson weights, and positive Pearson weights. The Pearson coefficient is
\begin{equation}
 \rho_{ij}=\frac{\sum_{u\in\mathcal T_{ij}}[x_i(u)-\bar x_i][x_j(u)-\bar x_j]}
 {\sqrt{\sum_u[x_i(u)-\bar x_i]^2}\sqrt{\sum_u[x_j(u)-\bar x_j]^2}},
 \label{eq:pearson}
\end{equation}
where $x=k$ for nodes and $x=s$ for hyperedges. All sums are over the same common history $\mathcal T_{ij}$, and $\bar x_i,\bar x_j$ are the means on that history. Shifted Pearson uses $w_{ij}=1+\rho_{ij}$ and assigns $\rho_{ij}=0$ when the coefficient is undefined. Positive Pearson uses $w_{ij}=\rho_{ij}$ only for finite $\rho_{ij}>0$ and excludes all other candidates. Consequently, the shifted rule admits constant histories and the positive rule excludes them. Raw and log distance both use $\sigma=1$, with bandwidth interpreted in the corresponding history units. All comparisons use $\lambda_n=\lambda_e=2$, $\tau_n=\tau_e=80$, and either $m_n=m_e=1$ or $m_n=m_e=2$.

\begin{figure*}[!t]
\centering
\subfloat[$m_n=m_e=1$; node]{\includegraphics[width=.235\textwidth]{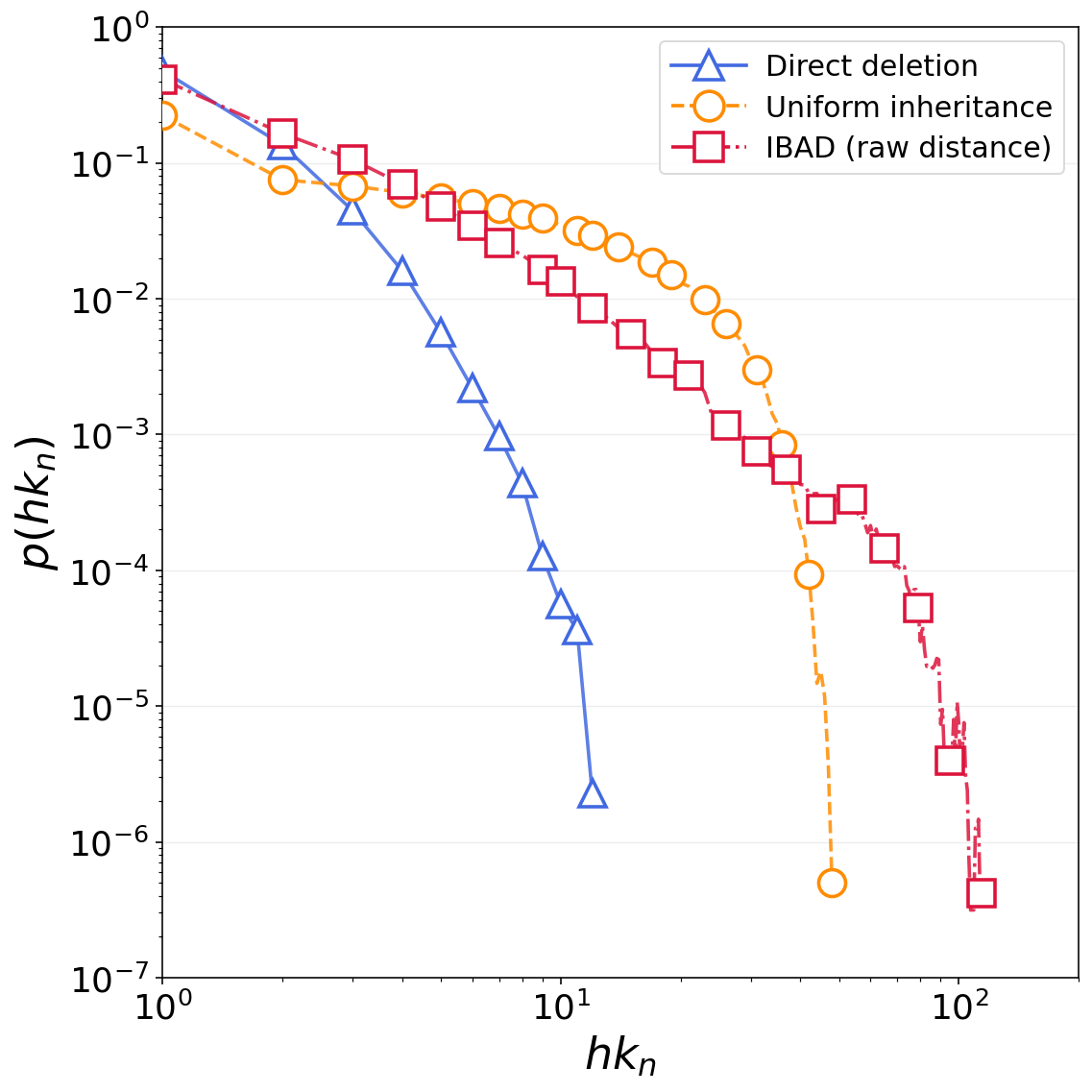}}\hfill
\subfloat[$m_n=m_e=1$; hyperedge]{\includegraphics[width=.235\textwidth]{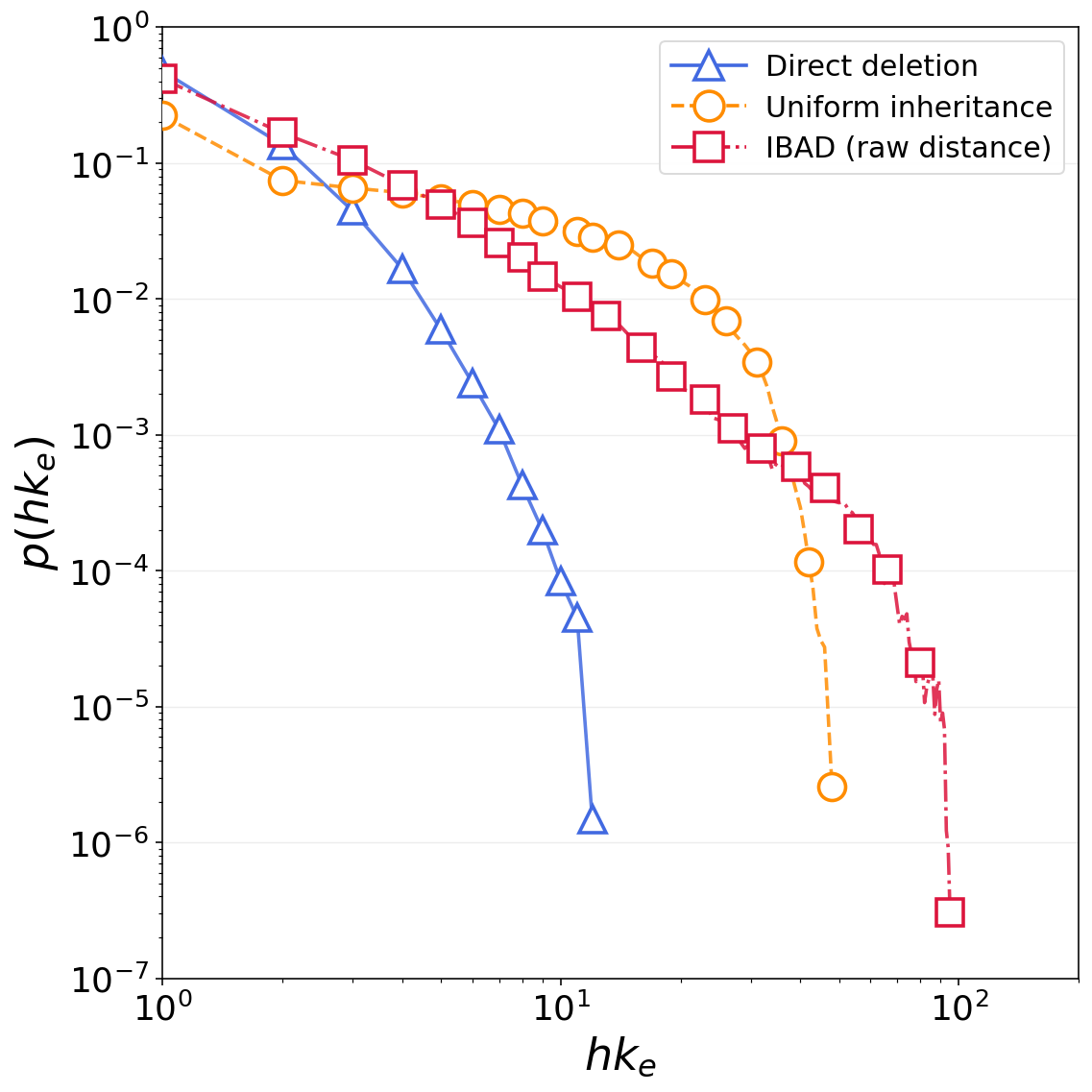}}\hfill
\subfloat[$m_n=m_e=2$; node]{\includegraphics[width=.235\textwidth]{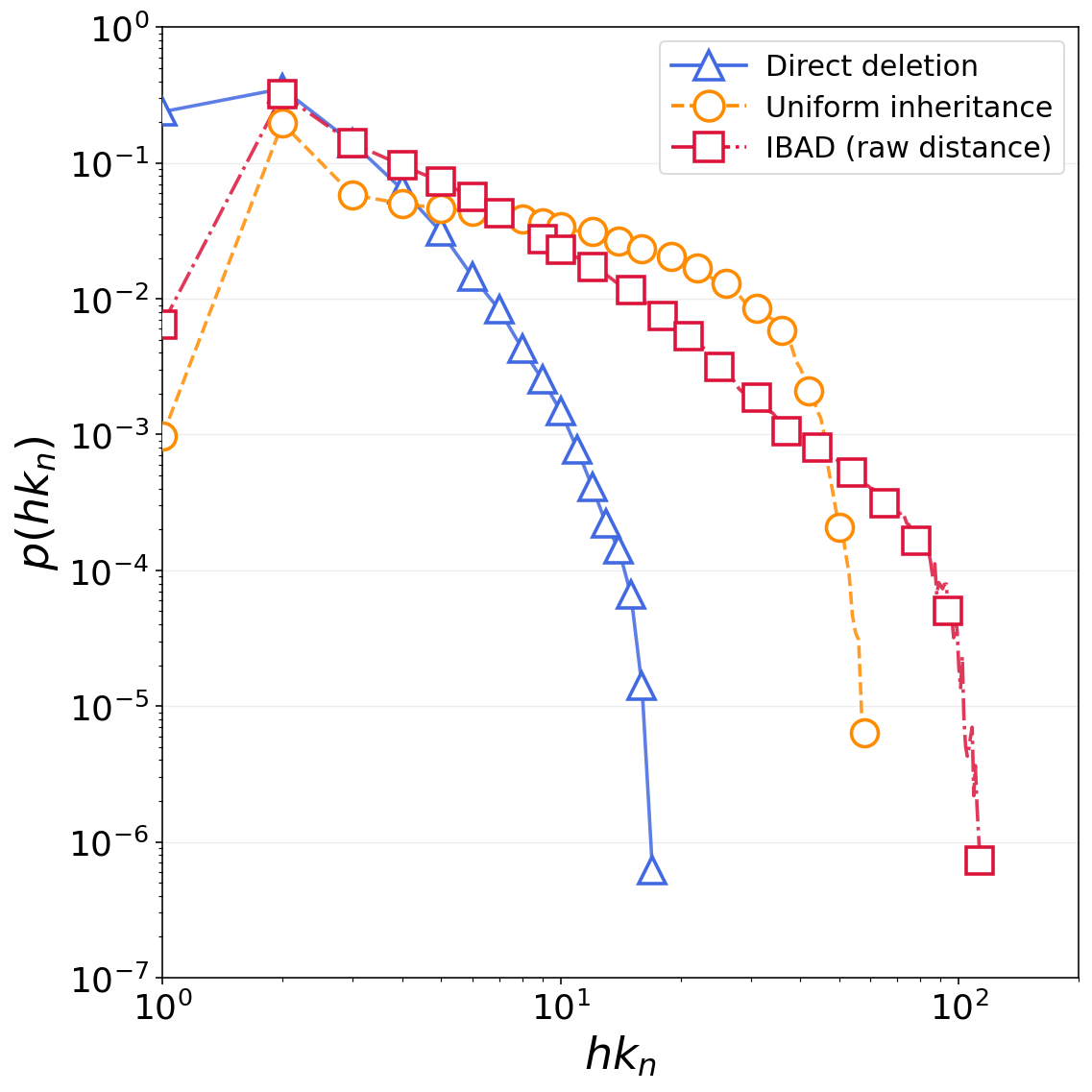}}\hfill
\subfloat[$m_n=m_e=2$; hyperedge]{\includegraphics[width=.235\textwidth]{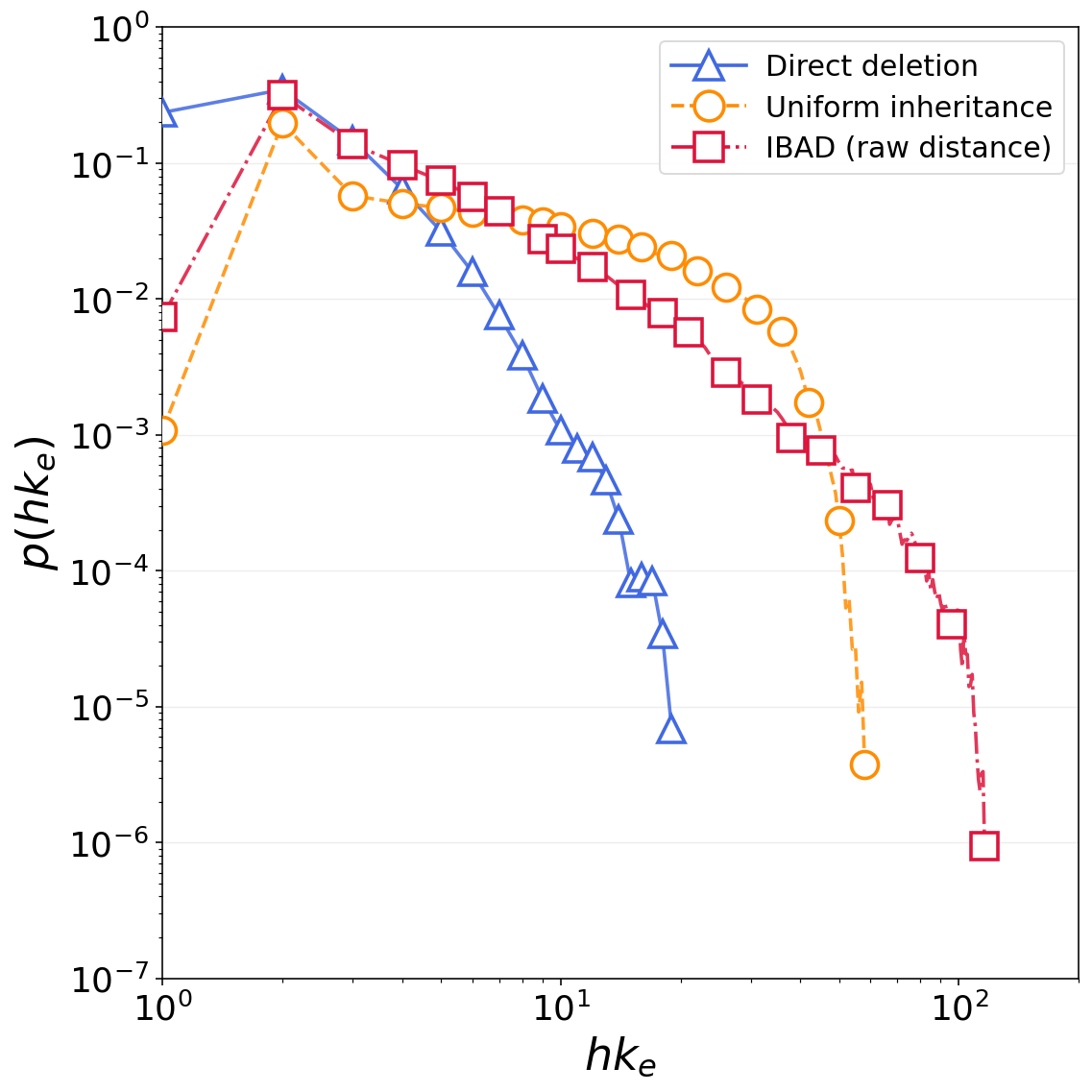}}\\[1pt]
\subfloat[$m_n=m_e=1$; node]{\includegraphics[width=.235\textwidth]{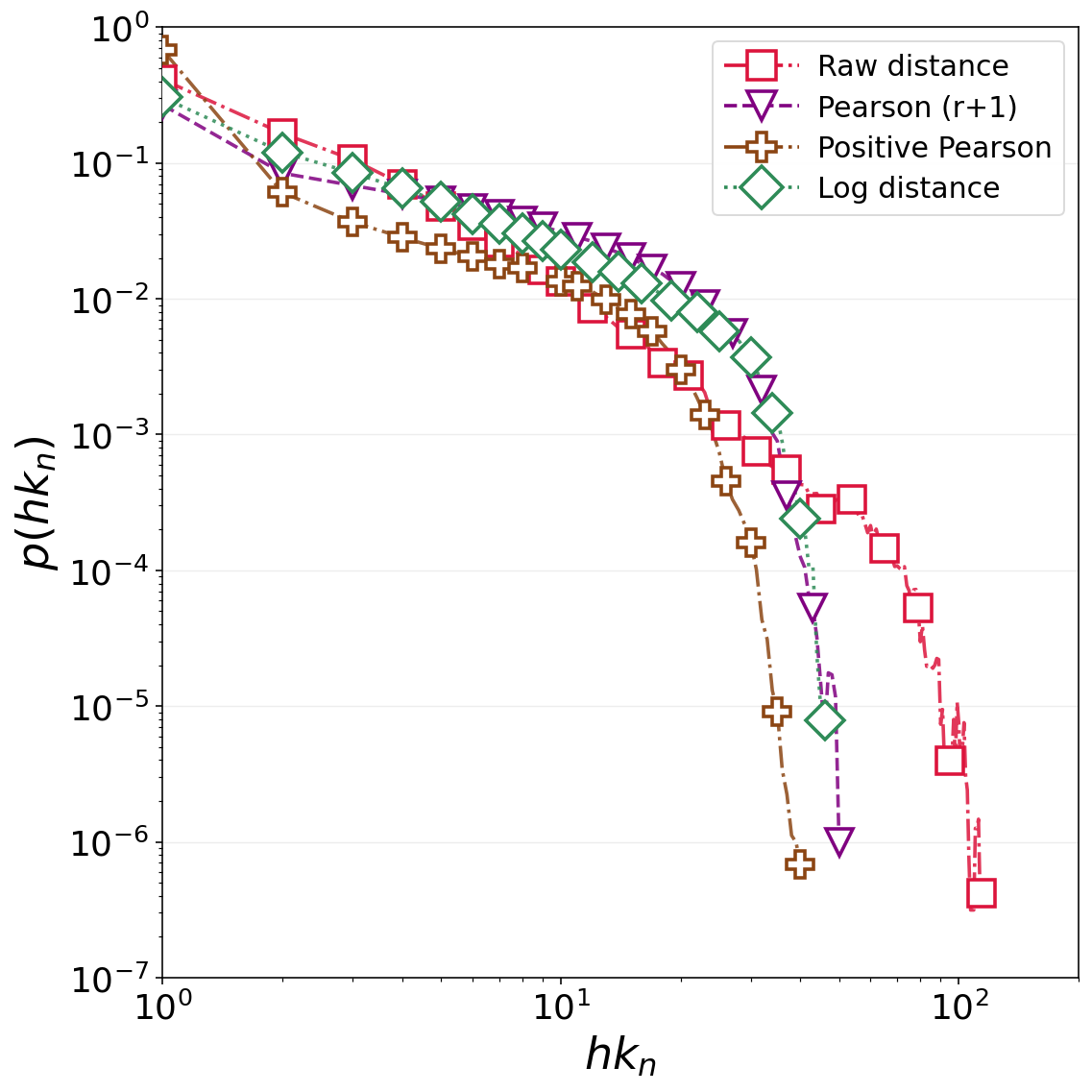}}\hfill
\subfloat[$m_n=m_e=1$; hyperedge]{\includegraphics[width=.235\textwidth]{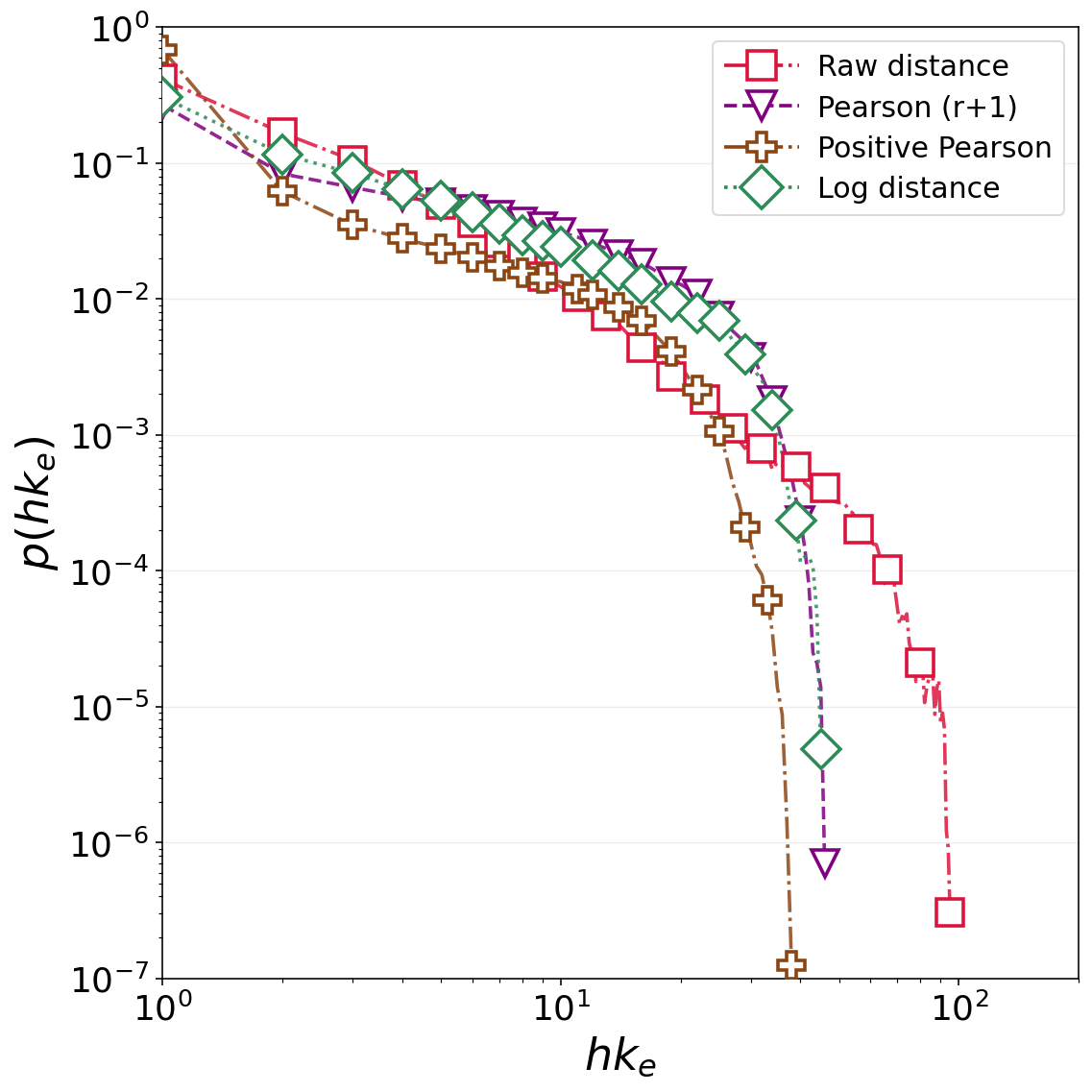}}\hfill
\subfloat[$m_n=m_e=2$; node]{\includegraphics[width=.235\textwidth]{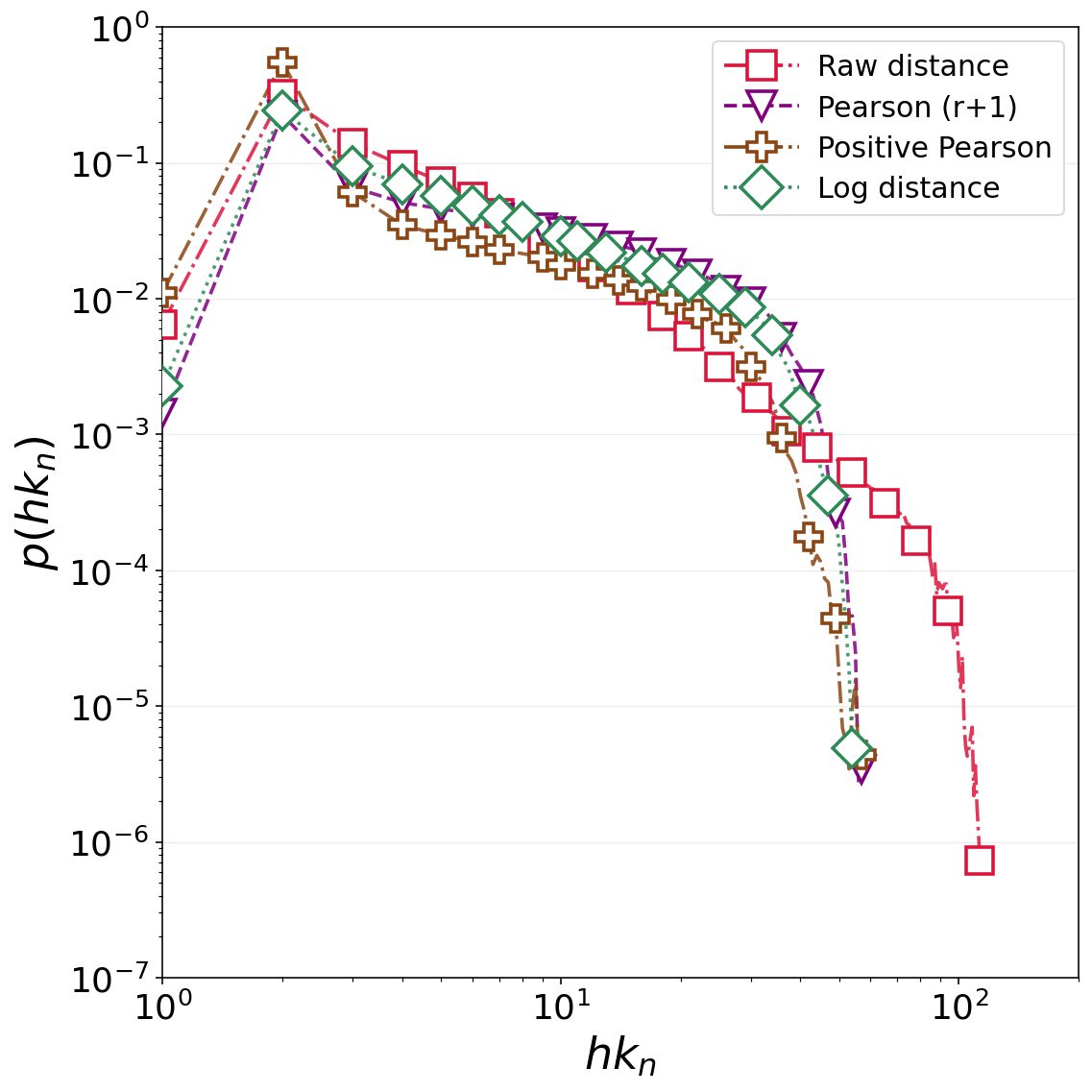}}\hfill
\subfloat[$m_n=m_e=2$; hyperedge]{\includegraphics[width=.235\textwidth]{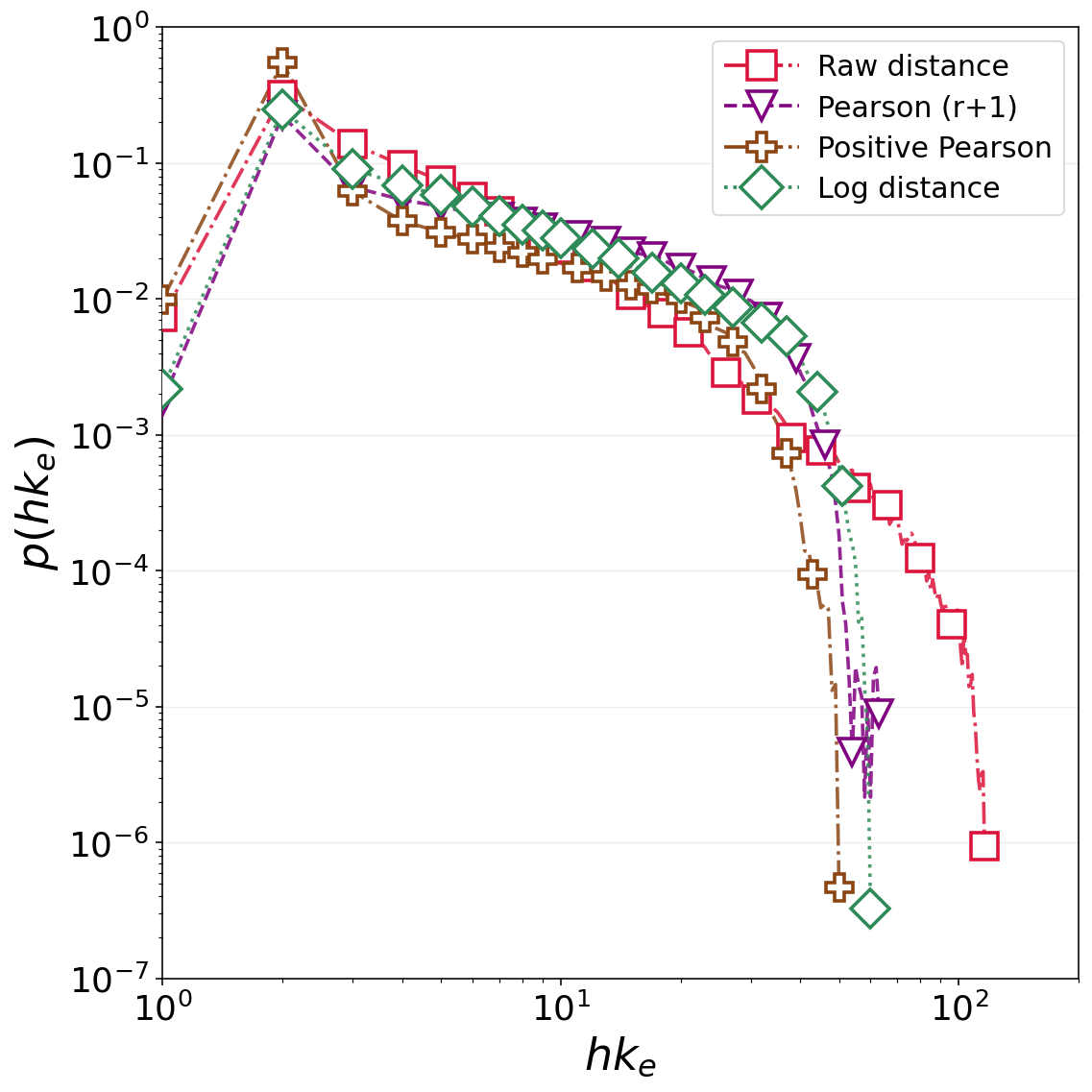}}\\[1pt]
\caption{\textbf{Node-degree and hyperedge-size distributions under different deletion mechanisms and similarity measures.} Panels (a)-(d) compare direct deletion, uniform single-proposal inheritance, and raw-history IBAD. Panels (e)-(h) compare raw distance, log distance, shifted Pearson, and positive Pearson under single-proposal inheritance. All use $\lambda_n=\lambda_e=2$ and $\tau_n=\tau_e=80$. Distance variants have numerical bandwidth one in their respective units. Both comparisons use the same raw-distance runs. Results are averaged over ten independent runs for each parameter setting, following the protocol at the beginning of Section~\ref{sec:simulations}.}\label{fig6}
\end{figure*}

\begin{table*}[!t]
\centering
\caption{Topology and Connection Retention Under Different Mechanisms}\label{tab:comparison}
\small
\setlength{\tabcolsep}{5pt}
\begin{tabular}{rlrrrrrrrr}
\hline
$m$ & Mechanism/similarity & $\langle k\rangle$ & $\langle s\rangle$ & ${\rm CV}_n$ & ${\rm CV}_e$ & $C$ & $R_{\rm LCC}$ & $\widehat\eta_n$ & $\widehat\eta_e$\\
\hline
1 & Direct & 1.004 & 1.008 & 0.991 & 1.002 & $0.0000\pm0.0000$ & 0.050 & -- & --\\
1 & Uniform & 8.203 & 8.331 & 0.929 & 0.930 & $0.4360\pm0.0056$ & 1.000 & 0.879 & 0.879\\
1 & Raw distance & 4.269 & 4.144 & 1.634 & 1.591 & $0.0936\pm0.0028$ & 0.997 & 0.764 & 0.763\\
1 & Shifted Pearson & 7.570 & 7.796 & 0.989 & 0.998 & $0.3988\pm0.0039$ & 0.999 & 0.868 & 0.870\\
1 & Positive Pearson & 3.068 & 3.088 & 1.418 & 1.441 & $0.0795\pm0.0024$ & 0.997 & 0.672 & 0.676\\
1 & Log distance & 6.565 & 6.551 & 1.129 & 1.123 & $0.3079\pm0.0031$ & 0.999 & 0.849 & 0.847\\
2 & Direct & 2.017 & 2.025 & 0.777 & 0.782 & $0.0211\pm0.0005$ & 0.815 & -- & --\\
2 & Uniform & 12.158 & 12.078 & 0.840 & 0.838 & $0.5074\pm0.0033$ & 1.000 & 0.838 & 0.838\\
2 & Raw distance & 7.258 & 7.235 & 1.340 & 1.315 & $0.2027\pm0.0035$ & 1.000 & 0.724 & 0.720\\
2 & Shifted Pearson & 11.680 & 11.555 & 0.883 & 0.882 & $0.4723\pm0.0012$ & 1.000 & 0.830 & 0.829\\
2 & Positive Pearson & 6.302 & 6.296 & 1.178 & 1.159 & $0.2211\pm0.0044$ & 1.000 & 0.685 & 0.681\\
2 & Log distance & 10.001 & 10.586 & 0.948 & 0.993 & $0.3999\pm0.0058$ & 1.000 & 0.802 & 0.808\\
\hline
\end{tabular}
\par\smallskip\begin{minipage}{0.97\textwidth}\footnotesize $m_n=m_e=m$, $\lambda_n=\lambda_e=2$, and $\tau_n=\tau_e=80$. $C$ is mean $\pm$ sample standard deviation of independent-run means. ${\rm CV}$ is the standard deviation divided by the mean of the averaged degree/size distribution. $\widehat\eta$ is the pooled total number of successful replacements divided by the total number of connections removed at deletion; direct deletion makes no proposals. $R_{\rm LCC}$ is averaged over sampled states and rounded to three decimals.\end{minipage}
\end{table*}

Figure~\ref{fig6} and Table~\ref{tab:comparison} compare the degree distributions and structural properties obtained under different deletion mechanisms. At $m_n=m_e=1$, direct deletion produces a fragmented network, with $R_{\rm LCC}\simeq0.050$ and $C=0$, whereas all tested inheritance rules yield $R_{\rm LCC}\gtrsim0.997$. Inheritance thus helps maintain a large connected component by replacing connections lost during deletion. The resulting connection counts and clustering levels nevertheless differ substantially between inheritance rules. Raw-history IBAD gives a mean node degree of approximately 4.27 and $C\simeq0.094$, compared with 8.20 and 0.436 for uniform inheritance. At $m_n=m_e=2$, the corresponding mean degrees are 7.26 and 12.16, and the clustering coefficients are 0.203 and 0.507.

These differences can be understood through the number and placement of successful replacement connections. Uniform inheritance samples eligible recipients with equal probability, whereas raw-history inheritance favors individuals with similar degree histories. Such recipients may already share connections with the departing individual, causing some proposals to be rejected. The measured retention fractions are consistent with this explanation: at $m_n=m_e=1$, successful replacements during node deletion account for approximately 0.764 of the removed connections under raw-history inheritance, compared with 0.879 under uniform inheritance. Consistent with \eqref{eq:L-drift}, lower retention under raw-history inheritance accompanies its smaller mean degree.

Shifted Pearson and log-distance inheritance produce larger mean degrees and clustering coefficients than raw-history inheritance in both tested settings. Shifted Pearson admits weakly correlated and constant histories, while the logarithmic transformation compresses absolute degree differences, especially at large degrees. Raw-history similarity remains sensitive to these differences and therefore favors recipients with closer absolute connectivity levels. Positive Pearson additionally excludes candidates with undefined or nonpositive correlations, reducing the opportunities for inheritance. Its clustering is below that of raw-history inheritance at $m_n=m_e=1$ and above it at $m_n=m_e=2$, indicating that the effect of candidate restriction depends on the attachment setting.

In the tested settings, raw-history IBAD maintains a large connected component with fewer connections and lower clustering than uniform inheritance. Its degree coefficient of variation is also larger than those obtained with uniform, shifted Pearson, and log-distance inheritance, indicating greater relative degree heterogeneity.

\subsection{Simulation Results for Real-World Network Modeling}

The three real-world networks are a plant-pollinator network (PPN) \cite{Petanidou1991,weblife}, a food web (FW) \cite{Bascompte2005}, and an email hypergraph (EN) \cite{Benson2018,Yin2017,Leskovec2007,cornell}. The matrix datasets are binarized at positive entries, retaining all rows and columns. For PPN, columns represent nodes, and each row defines a hyperedge containing the column species with positive entries, giving 666 nodes and 131 hyperedges. For FW, the incidence representation uses rows as nodes and columns as hyperedges. Each column defines a hyperedge containing the row species with positive entries, giving 244 nodes and 248 hyperedges. For EN, consecutive node identifiers are grouped according to the supplied hyperedge-size records, with repeated identifiers within a record collapsed. The present comparison uses the first 470 hyperedge records in file order, approximately 0.2\% of the 235,263 records, comprising 374 distinct nodes. Repeated hyperedge records remain separate interactions. 

Table~\ref{tab:table4} reports the simulation parameters. Each run starts from ten nodes and one hyperedge for PPN and EN, and from one node and ten hyperedges for FW. For the birth rates in Table~\ref{tab:table4}, mean lifetimes set the stationary population targets. For PPN these targets are $N^*=666$ and $M^*=130$, compared with the observed 666 and 131; for FW they are 244 and 248, and for EN 374 and 470. Connectivity and degree distributions are evaluated separately in the comparisons below.

\begin{table*}[t]
\centering
\caption{Empirical Sizes, Simulation Parameters, and Distribution Distances}\label{tab:table4}\label{tab:real-divergence}
\small\setlength{\tabcolsep}{6pt}\renewcommand{\arraystretch}{1.15}
\begin{tabular}{@{}lrrrrrrrrrrrr@{}}
\toprule
\multirow{2}{*}{Network} & \multirow{2}{*}{$N_{\rm real}$} & \multirow{2}{*}{$M_{\rm real}$} & \multirow{2}{*}{$\lambda_n$} & \multirow{2}{*}{$\lambda_e$} & \multirow{2}{*}{$\tau_n$} & \multirow{2}{*}{$\tau_e$} & \multirow{2}{*}{$m_n$} & \multirow{2}{*}{$m_e$} & \multicolumn{2}{c}{Node degree} & \multicolumn{2}{c}{Hyperedge size}\\
\cmidrule(lr){10-11}\cmidrule(l){12-13}
 & & & & & & & & & $D_{\rm KL}^{(\epsilon)}$ & $D_{\rm JS}$ & $D_{\rm KL}^{(\epsilon)}$ & $D_{\rm JS}$\\
\midrule
PPN & 666 & 131 & 6 & 13 & 111 & 10 & 1 & 5 & 0.4711$^{\dagger}$ & 0.0510 & 2.0430$^{\dagger}$ & 0.2308\\
FW & 244 & 248 & 4 & 13 & 61 & $248/13$ & 1 & 3 & 0.5259$^{\dagger}$ & 0.0774 & 0.6557 & 0.1244\\
EN & 374 & 470 & 2 & 5 & 187 & 94 & 1 & 2 & 0.1728 & 0.0550 & 0.7272 & 0.1715\\
\bottomrule
\end{tabular}
\par\smallskip\begin{minipage}{.97\textwidth}\footnotesize $\tau_n=1/\mu_n$ and $\tau_e=1/\mu_e$ are mean lifetimes. All comparisons use raw-history IBAD. Divergences use natural logarithms, with $\epsilon=10^{-10}$ for the smoothed KL divergence. $\dagger$ denotes an infinite unsmoothed KL divergence owing to a zero simulated probability at an empirically observed degree.\end{minipage}
\end{table*}

The calibration routine ranks tested integer pairs $(m_n,m_e)$ by the equal-weight average of the node-degree and hyperedge-size JS divergences, divided by $\ln 2$, using event-averaged degree counts. It applies coarse screening followed by local integer refinement and longer multi-seed evaluation of shortlisted pairs; the selected pair is then evaluated with new random seeds.

\begin{figure*}[!t]
\centering
\includegraphics[width=.96\textwidth]{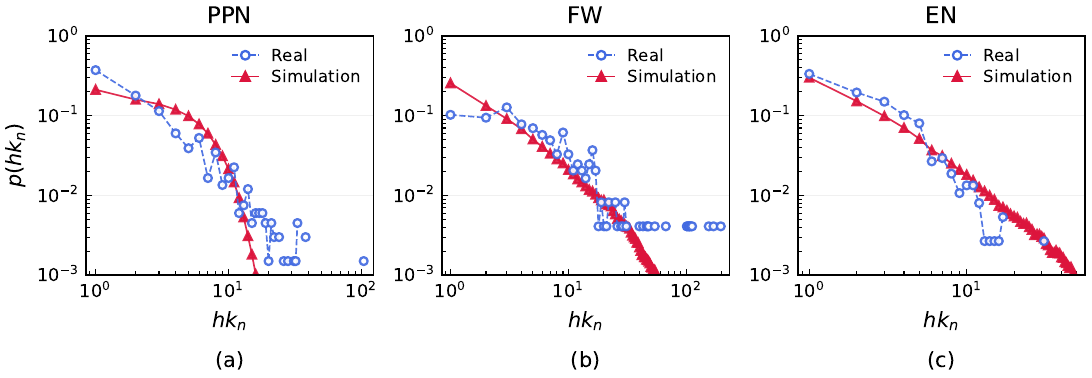}
\caption{\textbf{Empirical and simulated node-degree distributions.} Panels (a)-(c) compare PPN, FW, and EN, respectively. Blue circles denote empirical distributions and red triangles the distributions averaged from three simulations at the settings in Table~\ref{tab:table4}. Thin lines connect adjacent nonzero bins as visual guides. All panels use $10^{-3}\le p\le1$.}\label{fig8}
\end{figure*}

The empirical comparisons follow the burn-in and sampling protocol in Section~\ref{sec:simulations}. All degree bins, including zero degree and probabilities below the display limit in Fig.~\ref{fig8}, enter normalization and distribution-distance calculations.

We compare distributions on a common integer support, retaining zero-degree bins. Let $P$ denote the empirical distribution and $Q$ the averaged simulated distribution. With $U=(P+Q)/2$, the divergences are
\begin{align}
 D_{\rm KL}(P\Vert Q)&=\sum_{k:P_k>0}P_k\ln\frac{P_k}{Q_k},\label{eq:real-kl}\\
 D_{\rm JS}(P,Q)&=\tfrac12D_{\rm KL}(P\Vert U)+\tfrac12D_{\rm KL}(Q\Vert U).
 \label{eq:real-js}
\end{align}
Natural logarithms give a JS divergence in $[0,\ln 2]$. If $P_k>0$ but $Q_k=0$, the unsmoothed KL divergence is infinite. For a finite numerical summary we add $\epsilon=10^{-10}$ to every bin of each aligned vector and renormalize before evaluating $D_{\rm KL}^{(\epsilon)}$. JS uses the unsmoothed vectors. Table~\ref{tab:real-divergence} reports distances between the empirical distribution and the averaged simulated distribution.

Figure~\ref{fig8} compares the simulated and empirical node-degree distributions. Table~\ref{tab:real-divergence} reports distances for both sides. The node-degree JS divergences are 0.0510, 0.0774, and 0.0550 for PPN, FW, and EN, respectively, while the corresponding hyperedge-size divergences are 0.2308, 0.1244, and 0.1715. Under these representations, all three networks show closer distributional agreement on the node side.

\begin{table*}[!t]
\centering
\caption{Empirical and Simulated Topological Statistics}\label{tab:table5}
\small\setlength{\tabcolsep}{6pt}\renewcommand{\arraystretch}{1.10}
\begin{tabular}{@{}llrrrrrrrrr@{}}
\toprule
Network & Data & $N$ & $M$ & $L$ & $\langle k\rangle$ & $\langle s\rangle$ & $C$ & $R_{\rm LCC}$ & $\ell_{\rm LCC}$ & $Q_2$\\
\midrule
\multirow{2}{*}[-.5ex]{PPN} & Real & 666 & 131 & 2933 & 4.404 & 22.389 & 0.1439 & 0.9955 & 1.854 & 54100\\[3pt]
 & Simulation & \shortstack[r]{$669.6$\\[0pt]$\pm 7.3$} & \shortstack[r]{$129.8$\\[0pt]$\pm 0.7$} & \shortstack[r]{$2732.4$\\[0pt]$\pm 44.3$} & \shortstack[r]{$4.083$\\[0pt]$\pm 0.024$} & \shortstack[r]{$21.134$\\[0pt]$\pm 0.288$} & \shortstack[r]{$0.0548$\\[0pt]$\pm 0.0022$} & \shortstack[r]{$0.9979$\\[0pt]$\pm 0.0010$} & \shortstack[r]{$1.206$\\[0pt]$\pm 0.003$} & \shortstack[r]{$248375$\\[0pt]$\pm 9314$}\\
\midrule
\multirow{2}{*}[-.5ex]{FW} & Real & 244 & 248 & 3313 & 13.578 & 13.359 & 0.2563 & 1.0000 & 1.412 & 51457\\[3pt]
 & Simulation & \shortstack[r]{$243.8$\\[0pt]$\pm 5.9$} & \shortstack[r]{$247.7$\\[0pt]$\pm 1.8$} & \shortstack[r]{$2246.1$\\[0pt]$\pm 44.9$} & \shortstack[r]{$9.237$\\[0pt]$\pm 0.058$} & \shortstack[r]{$9.078$\\[0pt]$\pm 0.200$} & \shortstack[r]{$0.1130$\\[0pt]$\pm 0.0019$} & \shortstack[r]{$0.9992$\\[0pt]$\pm 0.0001$} & \shortstack[r]{$1.575$\\[0pt]$\pm 0.004$} & \shortstack[r]{$28952$\\[0pt]$\pm 1576$}\\
\midrule
\multirow{2}{*}[-.5ex]{EN} & Real & 374 & 470 & 1234 & 3.299 & 2.626 & 0.0164 & 0.8930 & 4.525 & 2648\\[3pt]
 & Simulation & \shortstack[r]{$360.3$\\[0pt]$\pm 2.4$} & \shortstack[r]{$474.7$\\[0pt]$\pm 4.6$} & \shortstack[r]{$3246.7$\\[0pt]$\pm 83.1$} & \shortstack[r]{$9.025$\\[0pt]$\pm 0.264$} & \shortstack[r]{$6.847$\\[0pt]$\pm 0.117$} & \shortstack[r]{$0.1066$\\[0pt]$\pm 0.0013$} & \shortstack[r]{$0.9997$\\[0pt]$\pm 0.0003$} & \shortstack[r]{$1.757$\\[0pt]$\pm 0.005$} & \shortstack[r]{$35503$\\[0pt]$\pm 1502$}\\
\bottomrule
\end{tabular}
\par\smallskip\begin{minipage}{.97\textwidth}\footnotesize Simulation entries are means with $\pm$ sample standard deviations of independent-run means on the second line. $\langle k\rangle$ and $\langle s\rangle$ average the snapshot ratios $L/N$ and $L/M$. $Q_2$ counts within-hyperedge node pairs with multiplicity.\end{minipage}
\end{table*}

Table~\ref{tab:table5} further compares their topological properties. In addition to clustering, connectivity, and path length, we report
\begin{equation}
	Q_2=\sum_{e\in E}\binom{s_e}{2},
	\label{eq:pair-incidences}
\end{equation}
which counts node pairs within hyperedges, including repeated contributions from pairs belonging to multiple hyperedges. For PPN, the simulated largest-component proportion remains close to the empirical value, while $C$ decreases from 0.1439 to 0.0548. The simulated network also has shorter largest-component paths and approximately 4.59 times the empirical $Q_2$. For FW, the mean node degree decreases from 13.578 to 9.237, and the mean connection count decreases from 3313 to approximately 2246. The clustering coefficient decreases from 0.2563 to 0.1130, while the largest-component proportion remains close to one. The mean path length within this component increases from 1.412 to 1.575, and $Q_2$ falls to approximately 56\% of the empirical count. Thus, the simulated FW remains largely connected but exhibits lower clustering and longer paths. These differences accompany a node-degree JS divergence of 0.0774. EN exhibits the opposite tendency: the mean node degree increases from 3.299 to 9.025, $R_{\rm LCC}$ rises from 0.8930 to 0.9997, and $C$ increases from 0.0164 to 0.1066. Its within-hyperedge node-pair count $Q_2$ reaches approximately 13.41 times the empirical value.

These structural differences can be understood partly through the contribution of large hyperedges. Since $Q_2=\tfrac12(\sum_e s_e^2-L)$, the node-pair count depends strongly on the second moment of hyperedge size. Consequently, similar mean hyperedge sizes can coexist with substantially different node-pair counts, as observed for PPN. Large hyperedges also connect many node pairs in the projection and can shorten paths. Their influence on the clustering coefficient additionally depends on connections across hyperedges, because triples contained within a single hyperedge are excluded from its calculation. Likewise, the relatively small node-degree JS divergence for EN is compatible with its larger mean degree: differences in low-probability, high-degree regions can contribute substantially to the mean while accounting for a limited fraction of the total probability.

The distributional comparisons and the reproduction of clustering, path lengths, and node-pair counts provide complementary assessments of the model.

\section{Conclusion}

This study proposed a higher-order dynamic network model governed by an inheritance-based deletion (IBAD) mechanism. The IBAD rule uses raw degree-history distance to propose one replacement for each connection removed at deletion. Existing connections cause rejection without another draw, so inheritance can leave a net loss of connections.

The theoretical analysis separates population stability from topological evolution. Node and hyperedge populations have Poisson stationary distributions determined by their arrival rates and mean lifetimes, regardless of the deletion rule. For IBAD, connection inheritance success probabilities, recipient gain distributions, and connection-count and degree drift relations explicitly show the roles of historical similarity and connection overlap.

The simulations exhibit broad heterogeneous degree distributions and approximate scaling over selected ranges. Birth rates and mean lifetimes mainly change population scale in the reported sweeps and have a comparatively weak effect on fitted slopes. Attachment parameters have coupled effects on distribution shape and clustering, with trends that depend on the parameter combination. Relative to direct deletion, inheritance markedly increases connectivity in the tested sparse settings. Relative to uniform inheritance, raw-history targeting retains fewer connections and yields lower clustering while preserving a large connected component. Alternative similarities change both heterogeneity and density, demonstrating that the similarity definition is a substantive modeling choice.

The empirical comparisons further delimit the model's scope. Node-degree agreement can coexist with mismatched hyperedge sizes, clustering, and within-hyperedge node-pair counts. Clustering is underestimated for PPN and FW, whereas EN clustering and connectivity are overestimated. These outcomes support evaluation against multiple distributional and structural criteria.

The current model remains limited in several respects. Similarity is measured only through degree history, which is a simplified assumption. Some deviations also remain in fitting empirical degree distributions. The bandwidth and history-eligibility rule require application-specific assessment. Statistical identification of power-law tails remains necessary beyond the descriptive log-log fits \cite{Clauset2009}. A closed stationary degree distribution and density-matched mechanism comparisons remain open problems. Future work may extend the model by incorporating weighted connections, directed hyperedges, or multilayer structures. It may also be useful to study the interaction between IBAD and other dynamic processes, such as diffusion and cascading failures, to better understand the coupling between network structure and function.

\ifCLASSOPTIONcaptionsoff
  \newpage
\fi

 \begin{IEEEbiography}[{\includegraphics[width=1in,height=1.25in,clip,keepaspectratio]{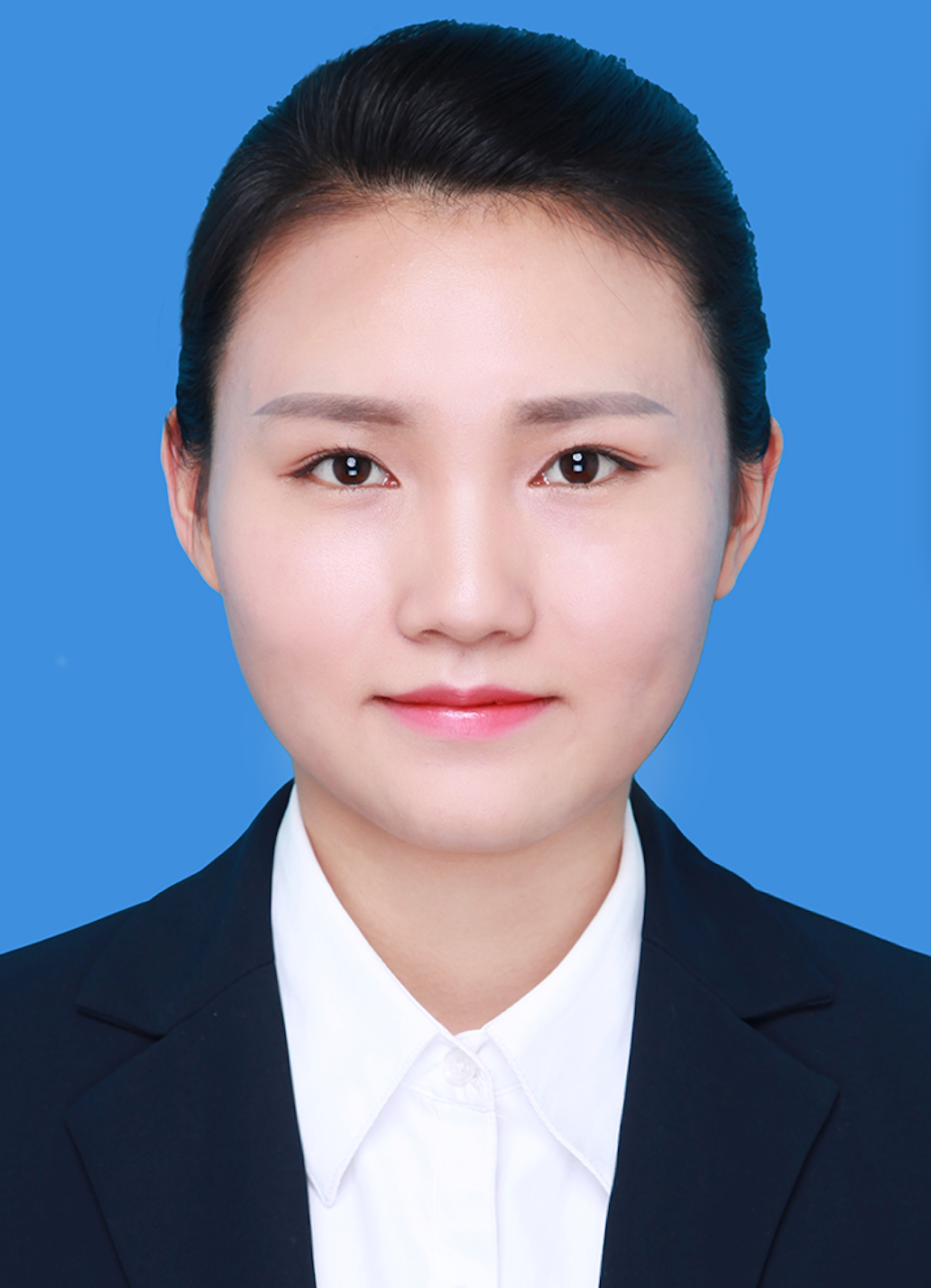}}]{Qin Li}
(Member, IEEE) received her Ph.D. degree in management from the School of Public Policy and Administration, Chongqing University, Chongqing, China, in 2025. Since 2025, she has been a lecturer in the Business College, Southwest University, Chongqing, China. Currently, she serves as an Editorial Board Member for \textit{Humanities \& Social Sciences Communications} and \textit{Scientific Reports}. Her research interests include Computational Social Science, Evolutionary Games, Epidemiological Models, and Complex Systems.
\end{IEEEbiography}

\begin{IEEEbiography}[{\includegraphics[width=1in,height=1.25in,clip,keepaspectratio]{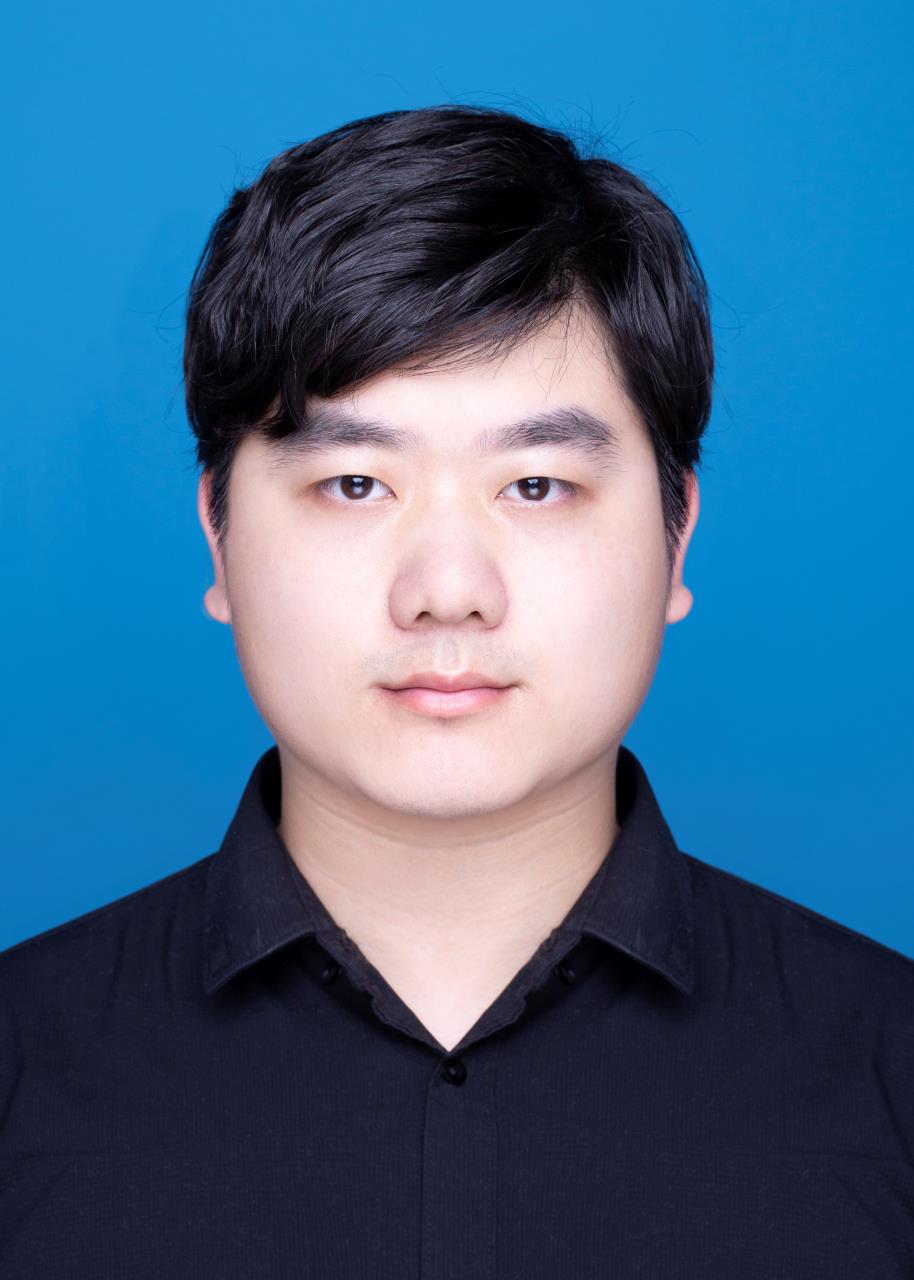}}]{Yichao Yao}
received the B.Eng.\ degree in data science and big data technology and the M.Eng.\ degree in computer science and technology from Southwest University, Chongqing, China, in 2024 and 2026, respectively. His research interests include complex networks, evolutionary games, stochastic processes, and nonlinear science.
\end{IEEEbiography}

\begin{IEEEbiography}[{\includegraphics[width=1in,height=1.25in,clip,keepaspectratio]{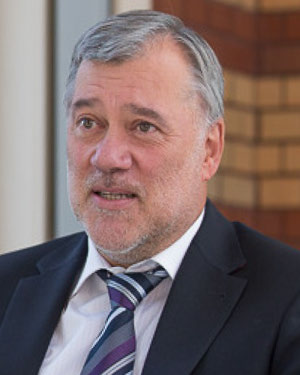}}]{J\"{u}rgen Kurths}
received the B.S. degree in mathematics from the University of Rostock, Rostock, Germany, the Ph.D. degree from the Academy of Sciences, German Democratic Republic, Berlin, Germany, in 1983, the Honorary degree from N. I. Lobachevsky State University, Nizhny Novgorod, Russia, in 2008, and the Honorary degree from Saratov State University, Saratov, Russia, in 2012.

From 1994 to 2008, he was a Full Professor with the University of Potsdam, Potsdam, Germany. Since 2008, he has been a Professor in nonlinear dynamics with the Humboldt University of Berlin, Berlin, Germany, and the Chair of the Research Domain Complexity Science with the Potsdam Institute for Climate Impact Research, Potsdam. He has authored more than 700 papers, which are cited more than 60\,000 times (H-index: 111). He is a Highly Cited Researcher in Engineering. His main research interests include synchronization, complex networks, and time series analysis and their applications. He is a member of the Academia Europaea. He is a Fellow of the American Physical Society, the Royal Society of Edinburgh, and the Network Science Society. He was a recipient of the Alexander von Humboldt Research Award from India in 2005 and from Poland in 2021, the Richardson Medal of the European Geophysical Union in 2013, and the Eight Honorary Doctorates. He is the Editor-in-Chief of \textit{Chaos} and on the editorial boards of more than ten journals.
\end{IEEEbiography}

\begin{IEEEbiography}[{\includegraphics[width=1in,height=1.25in,clip,keepaspectratio]{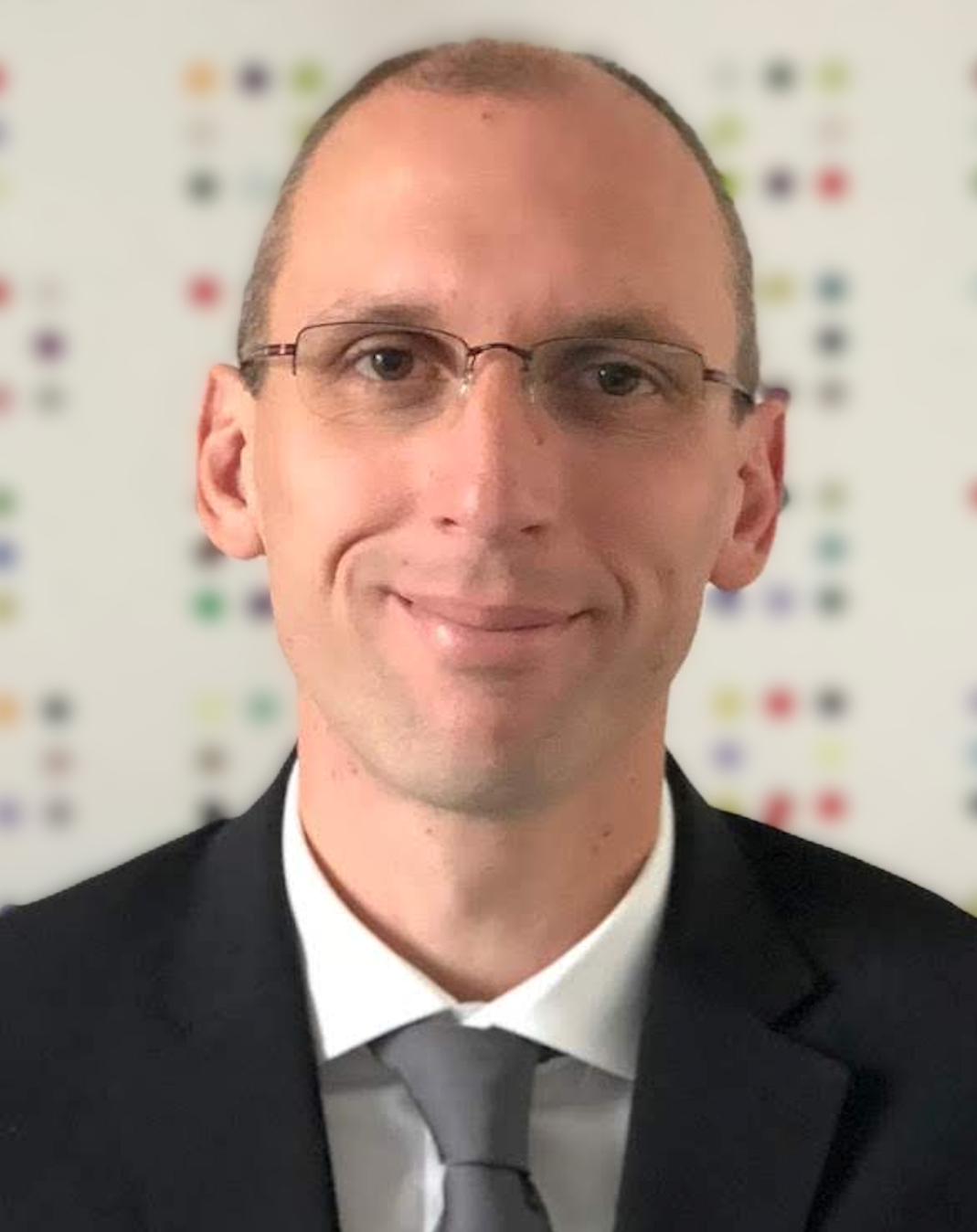}}]{Matja\v{z} Perc}
(Member, IEEE) received the Ph.D. degree in physics from the University of Maribor in 2006. He is currently Professor of Physics at the University of Maribor, Staff Researcher at the Community Healthcare Center Dr. Adolf Drolc Maribor, and Adjunct Professor at Kyung Hee University and Korea University. He is a member of Academia Europaea and the European Academy of Sciences and Arts, and among top 1\% most cited physicists according to Clarivate Analytics data. He is also the 2015 recipient of the Young Scientist Award for Socio and Econophysics from the German Physical Society, and the 2017 USERN Laureate. In 2018 he received the Zois Award, which is the highest national research award in Slovenia. In 2019 he became Fellow of the American Physical Society.
\end{IEEEbiography}

\begin{IEEEbiography}[{\includegraphics[width=1in,height=1.25in,clip,keepaspectratio]{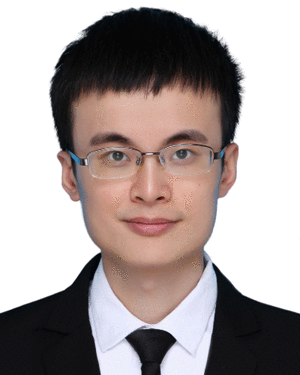}}]{Minyu Feng}
(Senior Member, IEEE) received his Ph.D. in Computer Science in 2018 through a joint program between the University of Electronic Science and Technology of China, Chengdu, China, and Humboldt University of Berlin, Berlin, Germany. Since 2019, he has been an Associate Professor at the College of Artificial Intelligence, Southwest University, Chongqing, China.

Dr. Feng has published more than 90 peer-reviewed papers in authoritative journals, including \textit{IEEE Transactions on Pattern Analysis and Machine Intelligence}, \textit{IEEE Transactions on Systems, Man, and Cybernetics: Systems}, and \textit{IEEE Transactions on Networking}. He is a Senior Member of the China Computer Federation (CCF) and the Chinese Association of Automation (CAA). Currently, he serves as a Subject Editor for \textit{Applied Mathematical Modelling} and an Editorial Advisory Board Member for \textit{Chaos}. He has also served as a current or former Editorial Board Member of more than 10 journals.

Dr. Feng's research interests include complex systems, evolutionary game theory, computational social science, and mathematical epidemiology.
\end{IEEEbiography}


\begin{thebibliography}{99}

\bibitem{barabasi2016}A.-L. Barab\'asi, \textit{Network Science}. Cambridge, U.K.: Cambridge Univ. Press, 2016.

\bibitem{newman2018}M. Newman, \textit{Networks}, 2nd ed. Oxford, U.K.: Oxford Univ. Press, 2018.

\bibitem{Bairey2016}E. Bairey, E. D. Kelsic, and R. Kishony, ``High-order species interactions shape ecosystem diversity,'' \textit{Nat. Commun.}, vol. 7, Art. no. 12285, 2016.

\bibitem{Grilli2017}J. Grilli, G. Barab\'as, M. J. Michalska-Smith, and S. Allesina, ``Higher-order interactions stabilize dynamics in competitive network models,'' \textit{Nature}, vol. 548, no. 7666, pp. 210-213, 2017.

\bibitem{Mastrandrea2015}R. Mastrandrea, J. Fournet, and A. Barrat, ``Contact patterns in a high school: A comparison between data collected using wearable sensors, contact diaries and friendship surveys,'' \textit{PLoS ONE}, vol. 10, no. 9, Art. no. e0136497, 2015.

\bibitem{Benson2016}A. R. Benson, D. F. Gleich, and J. Leskovec, ``Higher-order organization of complex networks,'' \textit{Science}, vol. 353, no. 6295, pp. 163-166, 2016.

\bibitem{Zhai2025}S. Zhai, J. Zhang, J. Ma, and Z. Xiang, ``Cluster synchronization of individuals during an epidemic: A contraction-based analysis,'' \textit{IEEE Trans. Syst., Man, Cybern., Syst.}, vol. 55, no. 10, pp. 6868-6878, 2025.

\bibitem{Li2024}X. Li, M. Chadli, Z. Tian, and W. Zhang, ``Resilient-learning control of cyber-physical systems against mixed-type network attacks,'' \textit{IEEE Trans. Syst., Man, Cybern., Syst.}, vol. 54, no. 9, pp. 5692-5703, 2024.

\bibitem{Bardoscia2021}M. Bardoscia \textit{et al.}, ``The physics of financial networks,'' \textit{Nat. Rev. Phys.}, vol. 3, no. 7, pp. 490-507, 2021.

\bibitem{Ando2022}T. Ando, M. Greenwood-Nimmo, and Y. Shin, ``Quantile connectedness: Modeling tail behavior in the topology of financial networks,'' \textit{Manag. Sci.}, vol. 68, no. 4, pp. 2401-2431, 2022.

\bibitem{majhi2022}S. Majhi, M. Perc, and D. Ghosh, ``Dynamics on higher-order networks: A review,'' \textit{J. Roy. Soc. Interface}, vol. 19, no. 188, Art. no. 20220043, 2022.

\bibitem{millan2025topology}A. P. Mill\'an \textit{et al.}, ``Topology shapes dynamics of higher-order networks,'' \textit{Nat. Phys.}, vol. 21, no. 3, pp. 353-361, 2025.

\bibitem{Feng2022}M. Feng, Y. Li, F. Chen, and J. Kurths, ``Heritable deleting strategies for birth and death evolving networks from a queueing system perspective,'' \textit{IEEE Trans. Syst., Man, Cybern., Syst.}, vol. 52, no. 10, pp. 6662-6673, 2022.

\bibitem{pi2025}B. Pi, L.-J. Deng, M. Feng, M. Perc, and J. Kurths, ``Dynamic evolution of complex networks: A reinforcement learning approach applying evolutionary games to community structure,'' \textit{IEEE Trans. Pattern Anal. Mach. Intell.}, vol. 47, no. 10, pp. 8563-8582, 2025.

\bibitem{feng2018}M. Feng, L. Deng, and J. Kurths, ``Evolving networks based on birth and death process regarding the scale stationarity,'' \textit{Chaos}, vol. 28, no. 8, Art. no. 083118, 2018.

\bibitem{zeng2023}Z. Zeng, M. Feng, and J. Kurths, ``Temporal network modeling with online and hidden vertices based on the birth and death process,'' \textit{Appl. Math. Model.}, vol. 122, pp. 151-166, 2023.

\bibitem{Feng2024}M. Feng, Z. Zeng, Q. Li, M. Perc, and J. Kurths, ``Information dynamics in evolving networks based on the birth-death process: Random drift and natural selection perspective,'' \textit{IEEE Trans. Syst., Man, Cybern., Syst.}, vol. 54, no. 8, pp. 5123-5136, 2024.

\bibitem{Zhang2025}H.-F. Zhang, K.-P. Liu, H. Wang, and C. Ma, ``Reconstructing network structures using Gaussian mixture model: From unsigned to signed networks,'' \textit{IEEE Trans. Syst., Man, Cybern., Syst.}, vol. 55, no. 11, pp. 8544-8557, 2025.

\bibitem{barabasi1999}A.-L. Barab\'asi and R. Albert, ``Emergence of scaling in random networks,'' \textit{Science}, vol. 286, no. 5439, pp. 509-512, 1999.

\bibitem{Li2003}X. Li and G. Chen, ``A local-world evolving network model,'' \textit{Physica A}, vol. 328, nos. 1-2, pp. 274-286, 2003.

\bibitem{Li2023}Y. Li, M. Feng, and J. Kurths, ``Evolving network modeling driven by the degree increase and decrease mechanism,'' \textit{IEEE Trans. Syst., Man, Cybern., Syst.}, vol. 53, no. 9, pp. 5369-5380, 2023.

\bibitem{Li2006}C. Li and G. Chen, ``Modelling of weighted evolving networks with community structures,'' \textit{Physica A}, vol. 370, no. 2, pp. 869-876, 2006.

\bibitem{Gu2008}Y. Gu and J. Sun, ``A local-world node deleting evolving network model,'' \textit{Phys. Lett. A}, vol. 372, no. 25, pp. 4564-4568, 2008.

\bibitem{Estrada2006}E. Estrada and J. A. Rodr\'iguez-Vel\'azquez, ``Subgraph centrality and clustering in complex hyper-networks,'' \textit{Physica A}, vol. 364, pp. 581-594, 2006.

\bibitem{Cencetti2021}G. Cencetti, F. Battiston, B. Lepri, and M. Karsai, ``Temporal properties of higher-order interactions in social networks,'' \textit{Sci. Rep.}, vol. 11, Art. no. 7028, 2021.

\bibitem{Battiston2020}F. Battiston \textit{et al.}, ``Networks beyond pairwise interactions: Structure and dynamics,'' \textit{Phys. Rep.}, vol. 874, pp. 1-92, 2020.

\bibitem{Battiston2021}F. Battiston \textit{et al.}, ``The physics of higher-order interactions in complex systems,'' \textit{Nat. Phys.}, vol. 17, no. 10, pp. 1093-1098, 2021.

\bibitem{Battiston2022}F. Battiston and G. Petri, Eds., \textit{Higher-Order Systems}. Cham, Switzerland: Springer, 2022.

\bibitem{bick2023}C. Bick, E. Gross, H. A. Harrington, and M. T. Schaub, ``What are higher-order networks?'' \textit{SIAM Rev.}, vol. 65, no. 3, pp. 686-731, 2023.

\bibitem{Bianconi2021}G. Bianconi, \textit{Higher-Order Networks}. Cambridge, U.K.: Cambridge Univ. Press, 2021.

\bibitem{Iacopini2019}I. Iacopini, G. Petri, A. Barrat, and V. Latora, ``Simplicial models of social contagion,'' \textit{Nat. Commun.}, vol. 10, Art. no. 2485, 2019.

\bibitem{St-Onge2021}G. St-Onge, H. Sun, A. Allard, L. H\'ebert-Dufresne, and G. Bianconi, ``Universal nonlinear infection kernel from heterogeneous exposure on higher-order networks,'' \textit{Phys. Rev. Lett.}, vol. 127, no. 15, Art. no. 158301, 2021.

\bibitem{Bodo2016}\'A. Bod\'o, G. Y. Katona, and P. L. Simon, ``SIS epidemic propagation on hypergraphs,'' \textit{Bull. Math. Biol.}, vol. 78, pp. 713-735, 2016.

\bibitem{Ferraz2023}G. F. de Arruda, G. Petri, P. M. Rodriguez, and Y. Moreno, ``Multistability, intermittency, and hybrid transitions in social contagion models on hypergraphs,'' \textit{Nat. Commun.}, vol. 14, Art. no. 1375, 2023.

\bibitem{Golubski2016}A. J. Golubski, E. E. Westlund, J. Vandermeer, and M. Pascual, ``Ecological networks over the edge: Hypergraph trait-mediated indirect interaction (TMII) structure,'' \textit{Trends Ecol. Evol.}, vol. 31, no. 5, pp. 344-354, 2016.

\bibitem{Levine2017}J. M. Levine, J. Bascompte, P. B. Adler, and S. Allesina, ``Beyond pairwise mechanisms of species coexistence in complex communities,'' \textit{Nature}, vol. 546, no. 7656, pp. 56-64, 2017.

\bibitem{Zlatic2009}V. Zlati\'c, G. Ghoshal, and G. Caldarelli, ``Hypergraph topological quantities for tagged social networks,'' \textit{Phys. Rev. E}, vol. 80, no. 3, Art. no. 036118, 2009.

\bibitem{Ferraz2021}G. F. de Arruda, M. Tizzani, and Y. Moreno, ``Phase transitions and stability of dynamical processes on hypergraphs,'' \textit{Commun. Phys.}, vol. 4, Art. no. 24, 2021.

\bibitem{Chen2024}L. Chen, Y. Xia, and Y. Zhang, ``Group synchronization of heterogeneous multiagent system with hierarchical structure and beyond pairwise interactions,'' \textit{IEEE Trans. Syst., Man, Cybern., Syst.}, vol. 55, no. 3, pp. 1578-1590, 2025.

\bibitem{Yuan2026}W. Yuan, W. Gu, T. Zhang, T. Tong, Q. Dong, and J. Sun, ``Hyperedge-based adaptive pinning control in hypernetworks under dynamic link attacks,'' \textit{IEEE Trans. Syst., Man, Cybern., Syst.}, vol. 56, no. 4, pp. 2474-2484, 2026.

\bibitem{Mulder2018}D. Mulder and G. Bianconi, ``Network geometry and complexity,'' \textit{J. Stat. Phys.}, vol. 173, no. 3, pp. 783-805, 2018.

\bibitem{young2017}J.-G. Young, G. Petri, F. Vaccarino, and A. Patania, ``Construction of and efficient sampling from the simplicial configuration model,'' \textit{Phys. Rev. E}, vol. 96, no. 3, Art. no. 032312, 2017.

\bibitem{petri2018}G. Petri and A. Barrat, ``Simplicial activity driven model,'' \textit{Phys. Rev. Lett.}, vol. 121, no. 22, Art. no. 228301, 2018.

\bibitem{Chodrow2020}P. S. Chodrow, ``Configuration models of random hypergraphs,'' \textit{J. Complex Netw.}, vol. 8, no. 3, Art. no. cnaa018, 2020.

\bibitem{Gallo2024}L. Gallo, L. Lacasa, V. Latora, and F. Battiston, ``Higher-order correlations reveal complex memory in temporal hypergraphs,'' \textit{Nat. Commun.}, vol. 15, Art. no. 4754, 2024.

\bibitem{Feng2024hyper}L. Feng \textit{et al.}, ``Hypernetwork modeling and topology of high-order interactions for complex systems,'' \textit{Proc. Natl. Acad. Sci. USA}, vol. 121, no. 40, Art. no. e2412220121, 2024.

\bibitem{Santoro2023}A. Santoro, F. Battiston, G. Petri, and E. Amico, ``Higher-order organization of multivariate time series,'' \textit{Nat. Phys.}, vol. 19, no. 2, pp. 221-229, 2023.

\bibitem{He2025}X. He, P. S. Chodrow, and P. J. Mucha, ``Edge correlations and link prediction in growing hypergraphs,'' \emph{Phys. Rev. E}, vol. 112, Art. no. 024305, 2025, doi: 10.1103/4lkd-mtzq.

\bibitem{Lerner2025}J. Lerner, M.-G. H\^{a}ncean, and M. Perc, ``Modeling temporal hypergraphs,'' \emph{J. Complex Netw.}, vol. 13, no. 6, Art. no. cnaf054, 2025, doi: 10.1093/comnet/cnaf054.

\bibitem{Jo2026}H.-H. Jo and N. Masuda, ``Modeling non-Poissonian temporal hypergraphs by Markovian node dynamics,'' \emph{Commun. Phys.}, vol. 9, Art. no. 337, 2026, doi: 10.1038/s42005-026-02885-0.

\bibitem{Wang2026reliability}F. Wang, N. Yu, and K. Li, ``A novel reliability assessment method for binary-state networks,'' \textit{Chaos Solitons Fractals}, vol. 205, Art. no. 117868, 2026.

\bibitem{Li2024virus}K. Li, Z. Chen, R. Cong, J. Zhang, and Z. Wei, ``Simulated dynamics of virus spreading on social networks with various topologies,'' \textit{Appl. Math. Comput.}, vol. 470, Art. no. 128580, 2024.

\bibitem{Dorogovtsev2000}S. N. Dorogovtsev, J. F. F. Mendes, and A. N. Samukhin, ``Structure of growing networks with preferential linking,'' \textit{Phys. Rev. Lett.}, vol. 85, pp. 4633-4636, 2000.

\bibitem{Rasmussen2006}C. E. Rasmussen and C. K. I. Williams, \textit{Gaussian Processes for Machine Learning}. Cambridge, MA, USA: MIT Press, 2006, ch. 4.

\bibitem{feng2017}M. Feng, H. Qu, Z. Yi, and J. Kurths, ``Subnormal distribution derived from evolving networks with variable elements,'' \textit{IEEE Trans. Cybern.}, vol. 48, no. 9, pp. 2556-2568, 2018.

\bibitem{Petanidou1991}T. Petanidou, ``Pollination ecology in a phryganic ecosystem,'' Ph.D. dissertation, Aristotle Univ. Thessaloniki, Thessaloniki, Greece, 1991.

\bibitem{weblife}``Web of Life: Ecological networks dataset.'' Accessed: Dec. 11, 2022. [Online]. Available: \url{https://www.web-of-life.es}

\bibitem{Bascompte2005}J. Bascompte, C. J. Meli\'an, and E. Sala, ``Interaction strength combinations and the overfishing of a marine food web,'' \textit{Proc. Natl. Acad. Sci. USA}, vol. 102, no. 15, pp. 5443-5447, 2005.

\bibitem{Benson2018}A. R. Benson, R. Abebe, M. T. Schaub, A. Jadbabaie, and J. Kleinberg, ``Simplicial closure and higher-order link prediction,'' \textit{Proc. Natl. Acad. Sci. USA}, vol. 115, no. 48, pp. E11221-E11230, 2018.

\bibitem{Yin2017}H. Yin, A. R. Benson, J. Leskovec, and D. F. Gleich, ``Local higher-order graph clustering,'' in \textit{Proc. 23rd ACM SIGKDD Int. Conf. Knowl. Discov. Data Mining}, 2017, pp. 555-564.

\bibitem{Leskovec2007}J. Leskovec, J. Kleinberg, and C. Faloutsos, ``Graph evolution: Densification and shrinking diameters,'' \textit{ACM Trans. Knowl. Discov. Data}, vol. 1, no. 1, Art. no. 2, 2007.

\bibitem{cornell}A. R. Benson, ``Datasets.'' Accessed: Dec. 11, 2022. [Online]. Available: \url{https://www.cs.cornell.edu/~arb/data/}

\bibitem{Clauset2009}A. Clauset, C. R. Shalizi, and M. E. J. Newman, ``Power-law distributions in empirical data,'' \textit{SIAM Rev.}, vol. 51, no. 4, pp. 661-703, 2009.

\end{thebibliography}
\end{document}